\documentclass[11pt]{article}
\usepackage{graphicx}
\usepackage{subfigure}

\usepackage{amsfonts}
\usepackage{amssymb}
\usepackage{amsmath}
\usepackage{geometry}
\usepackage{float}

\begin{document}


\title{Nonlinear shear wave interaction at a frictional interface: \\ Energy dissipation and generation of harmonics}
\author{A. Meziane \\  
Institut de M\'{e}canique et d'Ing\'{e}nierie de Bordeaux - I2M - UMR  CNRS 5295,\\ Universit\'{e} de Bordeaux, 351, Cours de la Lib\'{e}ration Talence 33405 Cedex, France.\\
A.N. Norris \\ 
Department of Mechanical and Aerospace Engineering, Rutgers University,  \\98 Brett Road, Piscataway, NJ 08854-8058, USA. \\ 
A.L. Shuvalov \\
Institut de M\'{e}canique et d'Ing\'{e}nierie de Bordeaux - I2M - UMR  CNRS 5295,\\ Universit\'{e} de Bordeaux, 351, Cours de la Lib\'{e}ration Talence 33405 Cedex, France. \\}
\date{}
\maketitle

\begin{abstract}

Analytical and numerical modelling of the nonlinear interaction of shear wave with a frictional interface is presented. The system studied is composed of two homogeneous and isotropic elastic solids,   brought into frictional contact   by remote normal compression. A shear wave, either time harmonic or a narrow band pulse, is incident normal to the interface and propagates through the contact. Two friction laws are considered and their influence on interface behavior is investigated : Coulomb's law  with a constant friction coefficient  and a  slip-weakening friction law which involves static and dynamic friction coefficients.  The relationship between the nonlinear harmonics and the dissipated energy, and their dependence on the contact dynamics (friction law, sliding and tangential stress) and on the normal contact stress are examined in detail. 
The analytical and numerical results indicate  universal type laws for the amplitude of the higher harmonics and for the dissipated energy, properly non-dimensionalized in terms of the pre-stress,  the friction coefficient and the incident amplitude.  
The results suggest that measurements of  higher harmonics can be used to quantify  friction and  dissipation effects of a {sliding interface}.

\end{abstract}

\section{Introduction}\label{sec1}

{The detection of  contacting interfaces which   can slide or open under applied load  is a difficult task but  
one that is relevant in many technologies.   Examples include 
closed cracks or defects, glued bondings, and partially delaminated material interfaces.     Ultrasonic methods based on linear wave scattering are efficient for detecting defects and   characterizing material elasticity, but are less sensitive to  closed cracks and contacting interfaces.} Using the nonlinear behavior of these defects, nonlinear ultrasonic techniques such as nonlinear resonance \cite{MNS3b}, sub- and higher-harmonic generation \cite{MNS2}, and frequency-modulation \cite{MNS1} have been shown to be sensitive to micro-cracks or closed cracks. For an overview of nonlinear acoustics applications see \cite{MNS3,MNS4,MNS5}. When an ultrasonic wave with large enough amplitude is incident on a contact with a frictional interface (e.g. closed cracks), higher harmonics appear in the frequency spectrum of transmitted and reflected waves. This effect, called Contact Acoustic Nonlinearity (CAN) \cite{MNS2}, is of increasing interest for  characterization of closed cracks or imperfectly bonded interfaces \cite{MNS3,MNS6}. 

Analysis of model systems  has helped to understand  the nonlinear interaction of waves and frictional cracks.  In particular, \cite{Chez83} analysed the partition of energy resulting from  a time harmonic obliquely incident plane SH wave  reflected and refracted by a frictional contact interface that slips.    Both longitudinal and shear wave propagation through a rough surface were investigated in \cite{MNS11} using an interface contact model    
based on Hertz theory, again for time harmonic incidence.  The partial contact model was   subsequently applied to model scattering from surface breaking cracks \cite{Pecorari05}, and numerical simulations indicated efficient production of second harmonics. Time domain studies have concentrated on numerical  implementations, such as   BEM formulation of SH slip motion on an arbitrary interface \cite{Mendelsohn89}.  Using a  generalization of this method   to include in-plane motion, it was shown in \cite{HA,Hirose94}  that the  amplitudes of the higher harmonics of the scattered far-fields can be useful in determining both the pre-stress and the frictional coefficient.  Measurements of second harmonic generation for normal incidence of longitudinal waves on a contacting interface between aluminum blocks have been reported in \cite{Biwa1,Biwa2}.  These experiments  indicate that the amplitude of the second harmonic  decreases rapidly with applied normal contact pressure initially, and then falls off in magnitude at a lesser rate.  
These findings are in agreement with experimental measurements on 
 contacting adhesive bonds \cite{MNS9}.  {A simple but instructive model was proposed in \cite{MNS10} for the nonlinear interaction of  an SH wave normally incident  on a frictional contact interface.  In contrast to the other studies mentioned, this  friction model predicts that only odd harmonics are generated for  time harmonic incident wave motion. 
}


{The objective here is to  investigate the effect of friction on shear wave scattering from  a interface. Physically the friction leads to nonlinearity due to switching between the sticking and sliding states of the interfacial contact. Analytical and numerical methods are combined  to understand the  role of energy loss and the generation of higher harmonics and their dependence  on the system parameters, such as applied normal load,  incident amplitude, and the friction coefficient.  We  consider a  shear} wave  incident normally on  a flat frictional interface. 
The friction behavior of the interface is described by two friction laws: Coulomb's law  with a constant friction coefficient  and a  slip-weakening friction law including static/dynamic friction coefficients.  
 {The outline of the paper is as follows.  The analysis in Section \ref{sec2} develops the basic model of \cite{MNS10} to consider both time harmonic and transient pulses.   Particular attention is given to estimating the energy lost in frictional sliding, and how it effects the scattered waves.  A closed form expression is obtained for the energy dissipation under  time harmonic incidence.} The time domain numerical model  is presented  in Section \ref{sec3}.  We note that the numerical simulation contains more of the actual physical effects encountered in practice, such as finite contact interface and cylindrical wave spreading. 
{Simulation } results  are discussed in Section \ref{sec4} in light of  the simpler analytical model, and several conclusions are drawn concerning the relation between {the total energy dissipated 
and the amplitudes of the  harmonics  in terms of non-dimensional system parameters.}

\section{{Nonlinear analysis of shear wave interaction}}\label{sec2}

\subsection{Setup and energy dissipation }\label{subsec1}

We consider uni-dimensional  SH motion in a solid with  shear modulus $G$,   density  $\rho$, impedance $Z=(G \rho)^{1/2}$ and 
wave speed  $c=(G/\rho)^{1/2}$.
A plane SH  wave  is normally  incident from $y>0$ on the  interface  $y=0$,  
such that the total shear stress $\sigma_{xy} \equiv \sigma_{\!_T}$ comprises the incident and reflected waves above the interface, and a transmitted wave below:
\begin{equation}\label{1}
\sigma_{\!_T} (t,y) = \begin{cases}
\sigma_{inc} + \sigma_{ref}, & y>0, 
\\
\sigma_{tra}, & y<0, 
\end{cases}
\quad
\text{where }\quad
\begin{matrix}
\sigma_{inc} = f(t+\frac{y}{c}),
\\
\sigma_{ref}= g(t-\frac{y}{c}),
\\
\sigma_{tra} = h(t+\frac{y}{c}).
\end{matrix}
\end{equation}
Instantaneous particle velocity in the $x$ direction is  
$v (t,y) $ and equal to  $Z^{-1}(\sigma_{inc} -  \sigma_{ref})$ for  $y>0$ and 
$Z^{-1}\sigma_{tra}$ for $y<0$.
Shear stress is  continuous across the interface, regardless of the state of contact, with   stress $ \sigma_{\!_T} (t,0+)=\sigma_{\!_T} (t,0-)
\equiv \sigma_{\!_{T0}} (t)$, where the latter follows from \eqref{1}  as
\begin{equation}\label{4}
\sigma_{\!_{T0}} (t) =f(t)+g(t) = h(t). 
\end{equation}
\begin{figure}[h]  
\begin{center} 
\subfigure{\includegraphics[width=2.9in, height=2.5in]{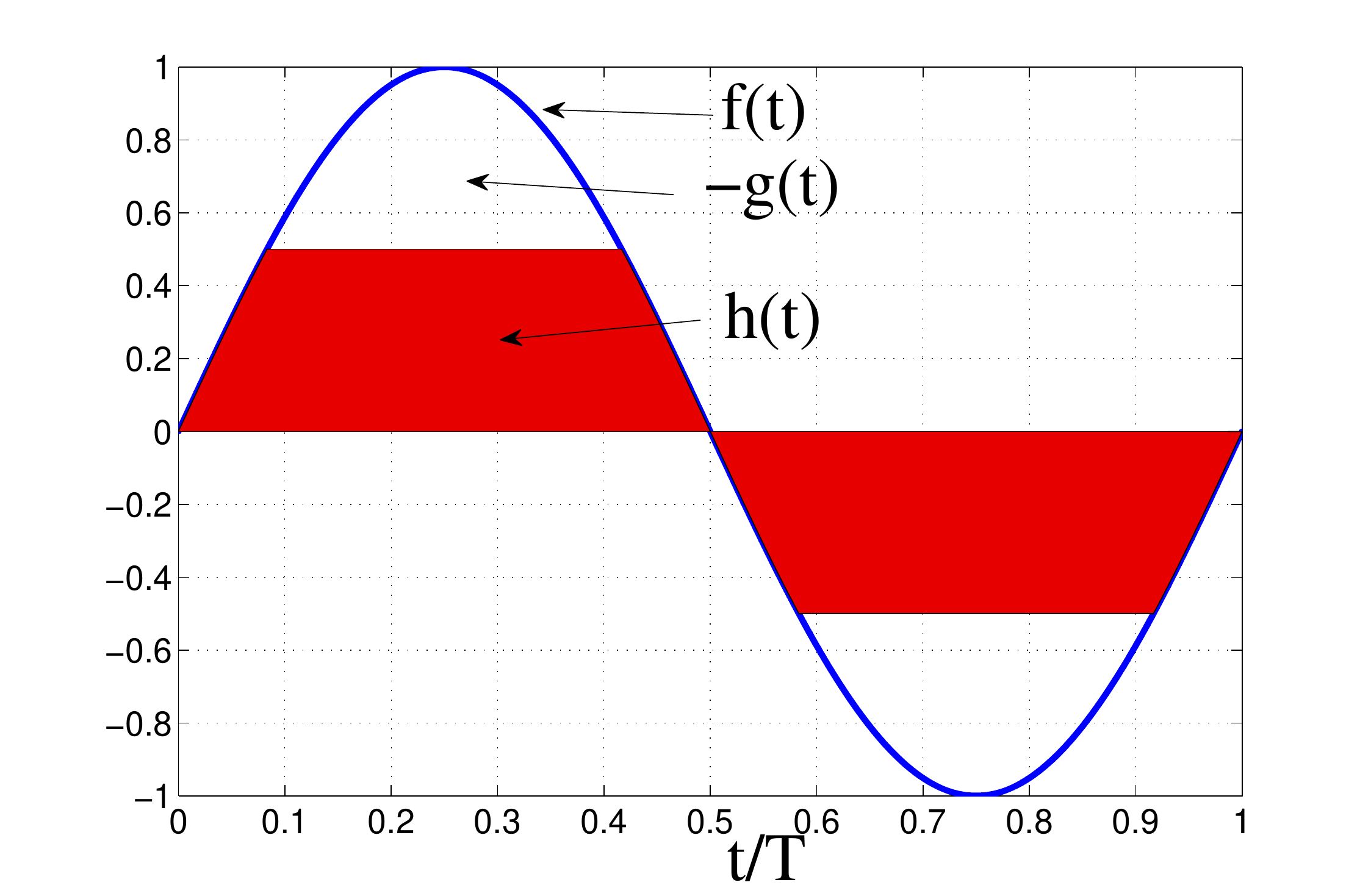}} 
\subfigure{\includegraphics[width=2.9in, height=2.5in]{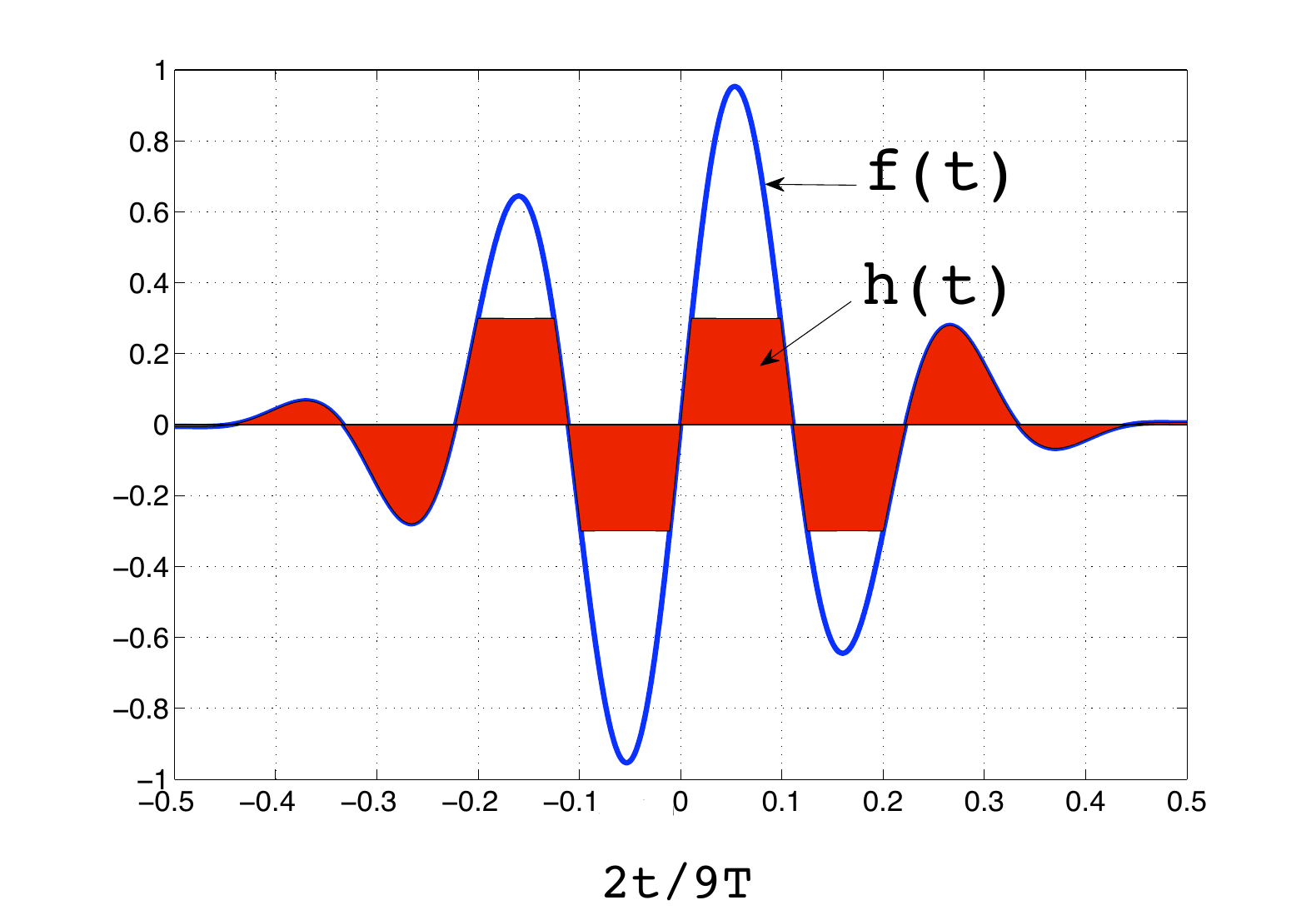}} 
\end{center} 
\caption{(a) The filled part depicts the transmitted wave for an incident time harmonic wave of period $T$ and  amplitude  twice the critical value $\sigma_{\!_{TC}} = 0.5$.  The remaining part of the wave form between $h$ and $f$ is the negative of the reflected wave $g(t)$, by Eq. \eqref{4}.
(b) A narrow band pulse for the incident  stress function  $f(t)$. The central frequency is $\omega = 2\pi/T$ with $T=\frac 29$,  $\sigma_{\!_{TC}}=0.3$.     }
 \label{fig2}
\end{figure}

When the interface is in contact the wave is completely transmitted with zero reflection:  $g(t) =0$,  $h(t) =f(t)$, and there is no energy dissipation.  
 When slipping occurs, the velocity jump across the interface 
follows from \eqref{1} - \eqref{4} as 
\begin{equation}\label{51}
[v (t)] \equiv v(t,0+)- v(t,0-) = -2 Z^{-1}g(t) . 
\end{equation}
{Slipping incurs irreversible  loss  of mechanical energy into heat, which can be quantified in two equivalent ways.  First, the work done per unit interface length by the slippage over a time increment $\Delta t$ is (force$\times$distance =) $\sigma_{\!_{T0}}   [v  ]\Delta t$.  The total 
 amount of energy dissipated  (per unit interface length) then follows from eqs. \eqref{4} and  
\eqref{51} as
\begin{equation}\label{7}
D  \equiv  \int \sigma_{\!_{T0}} (t) [v (t)] \, \text{d} t = -2 Z^{-1} \int   g (t)h(t)\, \text{d} t . 
\end{equation}
Alternatively, the rate of flux of incident energy per unit length (power flow) is (stress$\times$velocity =) $Z^{-1}\sigma_{inc}^2$.  The fluxes of the reflected and transmitted  waves are
 $Z^{-1}\sigma_{ref}^2$ and  $Z^{-1}\sigma_{tra}^2$, respectively.  The loss in energy (per unit length and time) is therefore $Z^{-1}(\sigma_{inc}^2-\sigma_{ref}^2-\sigma_{tra}^2)$. 
The three waves  are assumed to be plane waves, and hence the fluxes are independent of $y$, with $y> 0$ for the incident and reflected waves, and $y< 0$ for the transmitted wave.   Taking $y= \pm 0$ and using 
\eqref{1} to express the stresses, the total energy loss per unit length can be written  
\begin{equation}\label{new}
\text{wave energy lost} =  Z^{-1} \int   \big( f^2(t) - g^2 (t) -h^2(t)\big)\, \text{d} t  .
\end{equation}
Substituting from eq. \eqref{4} shows that the  lost wave energy is the same as the dissipated interfacial energy of eq. \eqref{7}.   
}

\begin{figure}[t]  
\begin{center} 
\subfigure[] {\includegraphics[width = 3.in, height=3.5in]{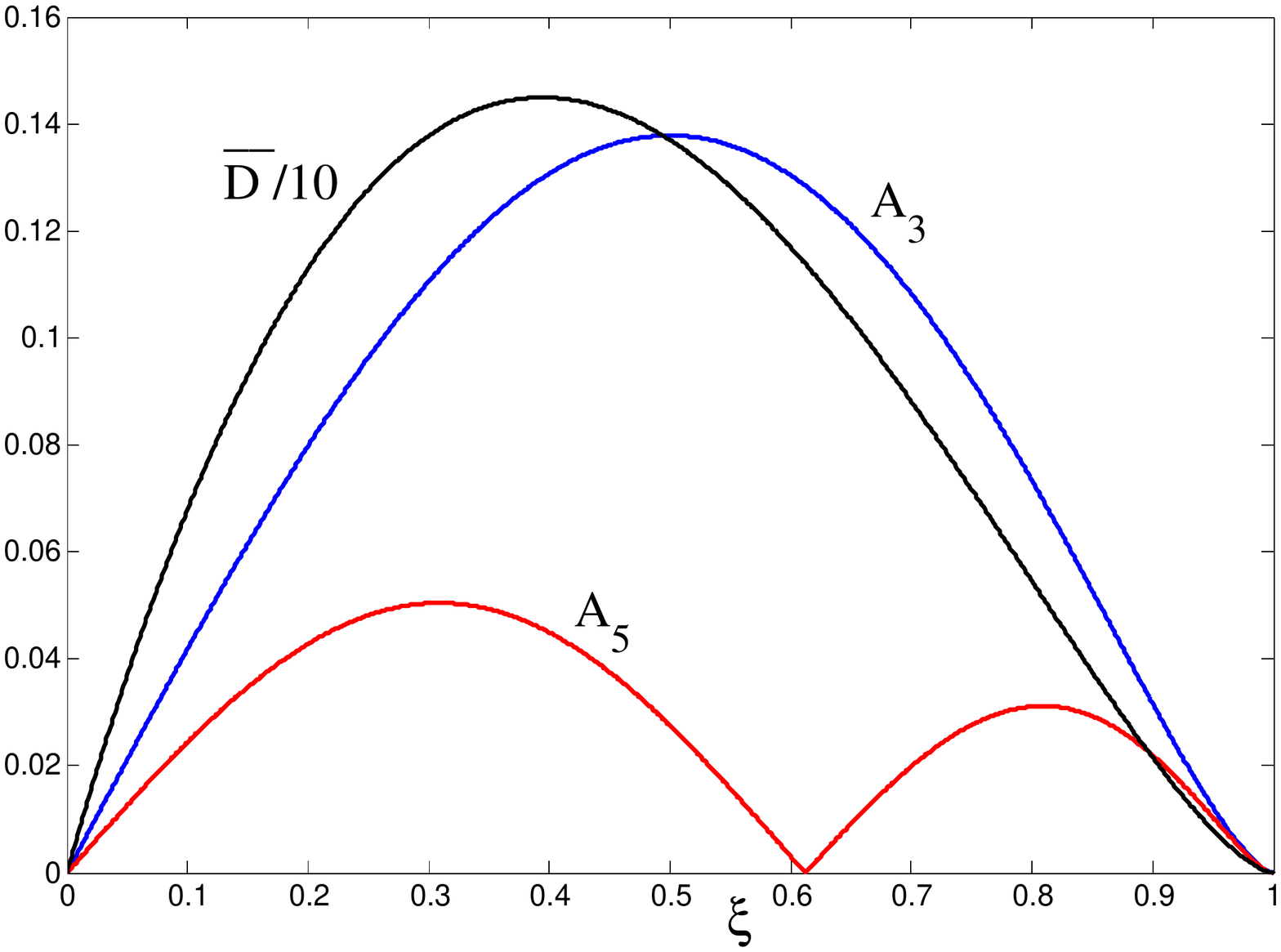}} 
\subfigure [ ]{\includegraphics[width = 3.in, height=3.5in]{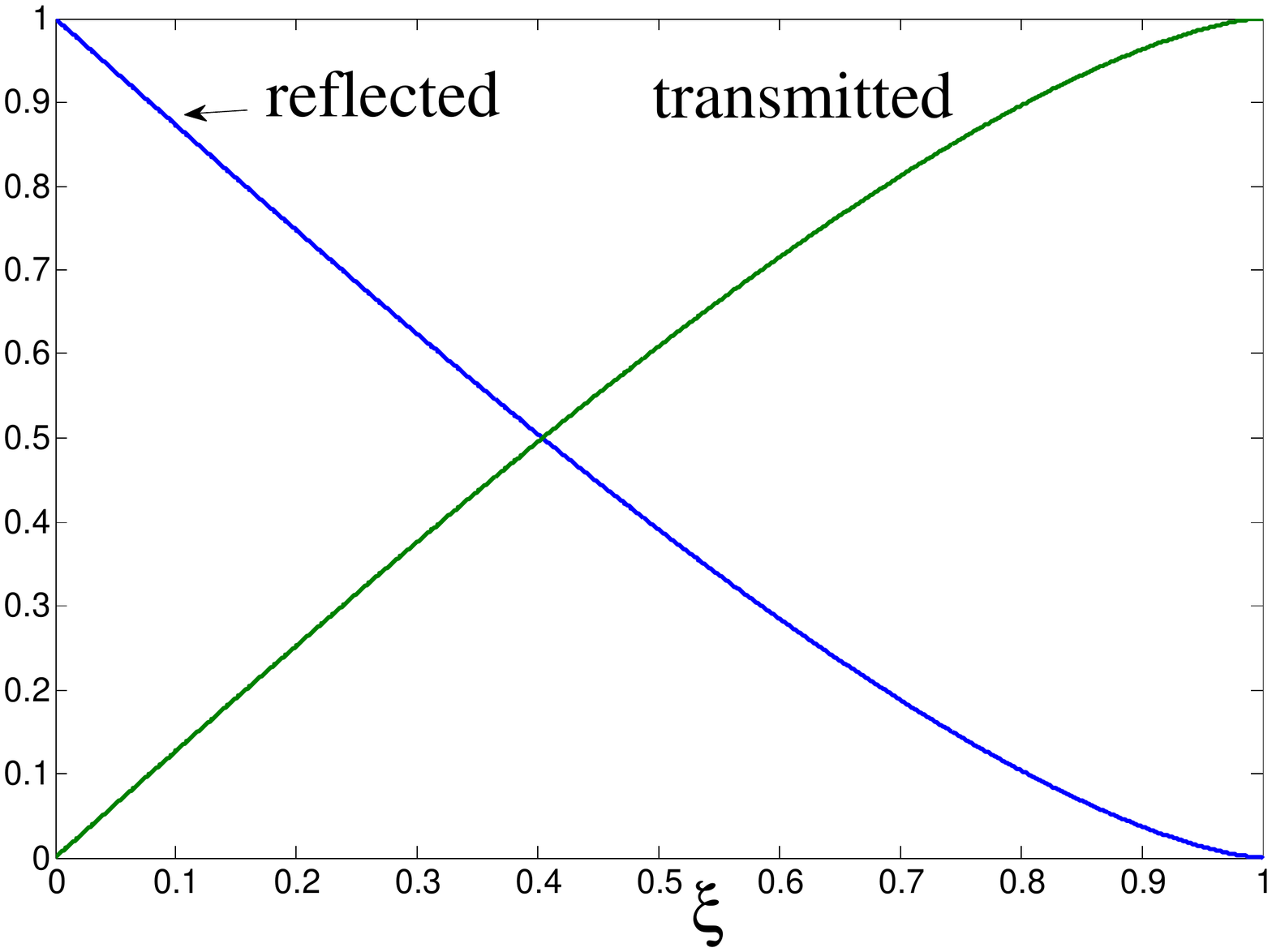}} 
\end{center} 
\caption{(a)  Magnitudes of the third and fifth harmonic amplitudes of the reflected or transmitted wave.  The third curve is
$\overline{D}/10$ where $\overline{D}$ is the nondimensional energy dissipation function  of \eqref{14}.   The abscissa  is $\xi= \mu |\sigma_{\!_N}|/\sigma_{\!_{TM}}=   \sigma_{\!_{TC}}/\sigma_{\!_{TM}}$, where  $\sigma_{\!_{TM}}$ is the maximum stress of the incident shear wave.   The maximum value of $A_3$ is at $\xi=0.5$, and  maximum  of $\overline{D}$ occurs at $\xi=0.3942$.
(b)  Reflected and   transmitted first harmonic amplitudes. 
}
 \label{fig3}
\end{figure}	

\subsection{Coulomb friction}

{Before the incident wave strikes the interface it is assumed to be in a state of } constant normal stress $\sigma_{\!_N} <0$ (compression). Slipping commences at the first instant that    
\begin{equation}\label{8}
|\sigma_{\!_{T0}}(t)|= \sigma_{\!_{TC}} ,  
\end{equation}
where the  critical value $\sigma_{\!_{TC}}$ depends on the friction model. 
 For simple 
\emph{Coulomb friction}: $ \sigma_{\!_{TC}} = \mu |\sigma_{\!_N}|$, where $\mu >0$ is the  
frictional coefficient.
	Slipping continues for  as long as 	$
	|f(t)|>  \sigma_{\!_{TC}}$ and  $|f(t)+g(t)|=  \sigma_{\!_{TC}}$ while it   occurs.
By considering the two possibilities: $f(t)>  \sigma_{\!_{TC}}$ ,  $f(t)<-  \sigma_{\!_{TC}}$,  
it follows that the reflected { wave during slipping is 
\begin{equation}\label{111}
g(t) = -\big( |f(t)| -  \sigma_{\!_{TC}} \big) \text{sgn}\big( f(t) \big)\, H( |f(t)| -  \sigma_{\!_{TC}}  ) 
, 
\end{equation}
where $H$ is the Heaviside step function.  The transmitted wave amplitude $h$ is given by eq. \eqref{4}
and the dissipated energy is  therefore } 
\begin{equation}\label{12}
D  = 2 Z^{-1}   \sigma_{\!_{TC}} \int\limits_{t: |f(t)|>  \sigma_{\!_{TC}}}   \big(|f(t)|- \sigma_{\!_{TC}} \big)\, \text{d} t . 
\end{equation}
{The slip condition \eqref{111} is equivalent to the  model proposed in \cite{MNS10}.}  It 
predicts that the reflected and transmitted waves are   ``clipped" versions of the incident pulse at the positive and negative values equal in magnitude to the critical yield stress, as illustrated in Figure \ref{fig2}.  {The simple expression \eqref{7} for the dissipated energy is new and will be central to subsequent developments.}

As an example,  consider the  time harmonic  incident wave   $f(t)= \sigma_{\!_{TM}} \sin \omega t$,  $\sigma_{\!_{TM}} > \sigma_{\!_{TC}}$.  	
	During a single period, $0\le t\le T=2\pi/\omega$ the slip condition $|f(t)|> \sigma_{\!_{TC}}$ prevails for $t\in (t_C,\frac{T}{2}-t_C)$ and $t\in (\frac{T}{2}+t_C,T-t_C)$ where 
	 $ \omega t_C = \theta_C \equiv     \sin^{-1} (\sigma_{\!_{TC}}/\sigma_{\!_{TM}} )\, 
	\in  (0,\pi /2)$, 
	with the same amount of energy lost in each sub-interval.  It follows that the reflected stress function is 
\begin{equation}\label{12=1}
-g(\omega^{-1}\theta) = \sigma_{\!_{TM}}
	(\sin \theta - \sin \theta_C)H( \theta - \theta_C) \ \ \text{for }\ 0<\theta < \pi/2.
\end{equation}
	The energy dissipated in one cycle   then  follows from Eq. \eqref{7} as  
\begin{equation}\label{14}
D    = \frac{   \sigma_{\!_{TM}}^2}{ \omega Z} 
\overline{D}  (\xi) ,
\quad
\xi 
= \frac{\sigma_{\!_{TC}}}{\sigma_{\!_{TM}}}, 
\quad
\overline{D} (\xi) = 
8 \xi \big[\big(1-\xi^2\big)^{1/2} - \xi
\big(\frac{\pi}{2} - \sin^{-1} \xi  \big) 
\big],
\end{equation}
where the nondimensional function $\overline{D}$ is shown in Figure \ref{fig3}.   Nonlinear effects occur only for values of the non-dimensional stress parameter $\xi$ less than unity.  If $\xi >1$  no energy is dissipated as the wave transmits unaltered.  We will also find the scaling factor $ { \omega Z}   \sigma_{\!_{TM}}^{-2}$ critical for interpreting the numerical simulations (see \S \ref{3.3}).

\subsection{Generation of higher  harmonics}

The reflected and transmitted stress waves together sum to equal the wave shape of the incident wave.  This means that the the higher harmonic amplitudes of the reflected and transmitted waves equal the content of the incident wave.  If the latter is a single frequency time harmonic wave, then higher harmonic amplitudes of the reflected and transmitted waves are equal in magnitude and of opposite sign.  We consider this case first. 

\subsubsection{Time harmonic wave motion}

The incident wave is  $f(t)= \sigma_{\!_{TM}} \sin \omega t$,  $\sigma_{\!_{TM}} > \sigma_{\!_{TC}}$.  The clipped nature of the reflected wave implies that  its Fourier series contains only odd harmonics
\cite{MNS10}, i.e. 
\begin{align}\label{21}
g(t) &= \sigma_{\!_{TM}}\, \sum\limits_{n=1}^\infty A_{2n-1} \sin ((2n-1)\omega t),
\quad
A_1(\xi) =   1 -\frac 2{\pi} \sin^{-1}\xi -\frac 2{\pi} \xi \big(1-\xi^2\big)^{1/2},
\nonumber \\
& A_3(\xi) =  -\frac 4{3\pi} \xi \big(1-\xi^2\big)^{3/2}, 
\quad
A_5(\xi) =  -\frac 4{15\pi} \xi (3-8 \xi^2) \big(1-\xi^2\big)^{3/2} ,  \ \ \text{etc.}, 
\end{align}
where  $\xi$ is defined in \eqref{14} .  
The first harmonic amplitudes and the magnitudes of the third and fifth harmonics are plotted in Figure \ref{fig3}. 
Note that by definition the non-dimensional amplitudes are normalized such  that $A_1+B_1=1$ where $B_1$ is the amplitude of the first harmonic of the transmitted wave. 
\begin{figure}[htb]  
\begin{center} 
\includegraphics[width=4.in,height=4.in,]{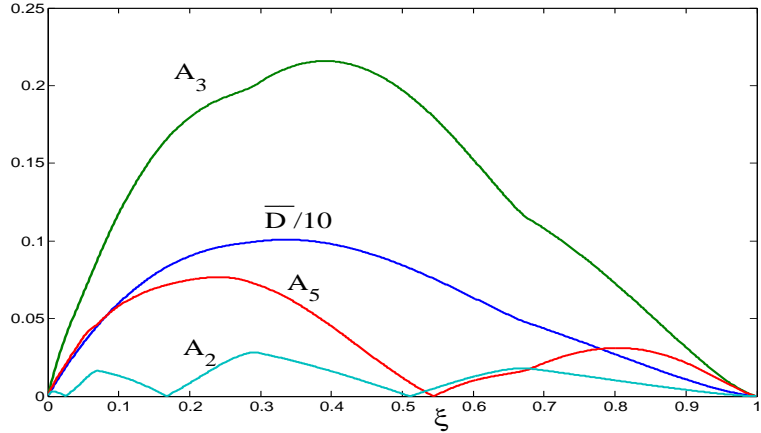}
\end{center} 
\caption{The dissipation and the second, third and fifth harmonic amplitudes as a function of 
$\xi= \sigma_{\!_{TC}}/\sigma_{\!_{TM}}$ where $\sigma_{\!_{TM}}$ is the amplitude of the incident wave in Figure \ref{fig2}b.}
 \label{fig5}
\end{figure}

\subsubsection{A narrow band pulse}

A narrow band pulse is shown in Figure \ref{fig2}b.  Note that the pulse is antisymmetric about the center $t=0$ because it represents the incident shear stress, which is proportional to the velocity.  The incident displacement amplitude is symmetric about $t=0$. 
Explicit expressions for the dissipation and for the harmonic content are not available, but are easily found by numerical quadrature.  Thus, the dissipation is computed from \eqref{12}, and the harmonic amplitudes of the reflected wave are given by analogy with \eqref{21}$_2$ as 
\begin{equation}\label{211}
A_n =  \frac{\omega }{\pi} \sigma_{\!_{TM}} \, \int \limits 
g(t) \sin (n\omega t) \text{d} t. 
\end{equation} 
Note that all of these quantities ($D$, $A_n$) are proportional to Fourier components of the reflected wave, either at zero frequency for $D$, or finite values for the harmonics. 
Figure \ref{fig5} shows the dissipation (normalized by the 
scaling factor $ { \omega Z}   \sigma_{\!_{TM}}^{-2}$ of Eq. \eqref{14}) and the magnitudes of the several  harmonics.  Comparing Figures \ref{fig3}a and \ref{fig5} we note that  the   harmonics $A_3$ and $A_5$ for the pulse are larger in magnitude as compared with the time harmonic case, whereas the  
nondimensional dissipation $\overline{D}$ is smaller.  The overall magnitudes are still comparable with those in the time harmonic case because, even though the integral is over all $t$  rather than a single period, the majority of the wave clipping and dissipation occurs within one cycle.
Note that $A_2 \ne 0$ but it is small in comparison with $A_3$.

We next consider a more realistic model using numerical simulation.

\section{Numerical study for  non-plane  wave incidence}\label{sec3}

\subsection{System studied}

\begin{figure}[htb]  
\begin{center} 
\subfigure[] {\includegraphics[width=3in, height=2.7in]{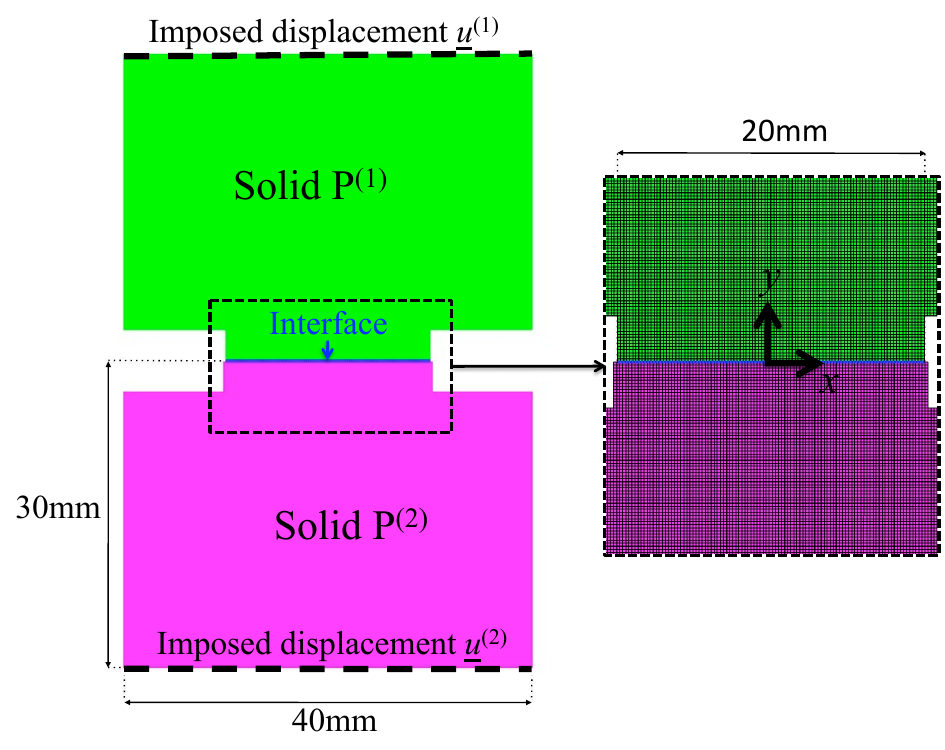}} 
\subfigure[] {\includegraphics[width=3in, height=2.7in]{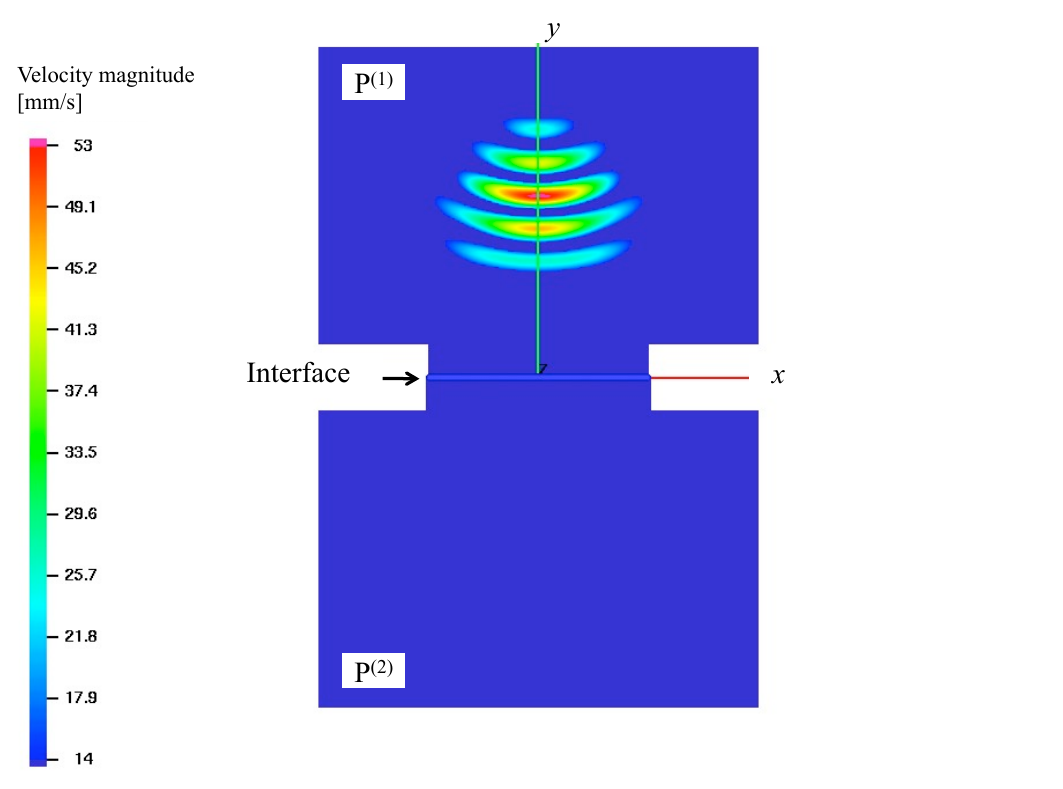}} 
\end{center} 
\caption{(a) The system considered in the  numerical model.
(b)    Snapshot of the velocity magnitude illustrating the non-plane wave incidence   generated from the  upper surface of P$^{(1)}$.   }
\label{fam1}
\end{figure}

The  system modeled is shown in Figure \ref{fam1}.  It is composed of two identical isotropic elastic media P$^{(1)}$ and P$^{(2)}$ separated by an interface governed by unilateral contact with a Coulomb friction law. The material used in the simulations is aluminium with the Young modulus, Poisson's ratio and density equal to 69GPa, 0.33 and 2700kg/m$^3$ respectively. The problem is considered in terms of a material (Lagrangian) description.

\begin{figure}[htb]  
\begin{center} 
\subfigure[] {\includegraphics[width = 3.in, height=3.in]{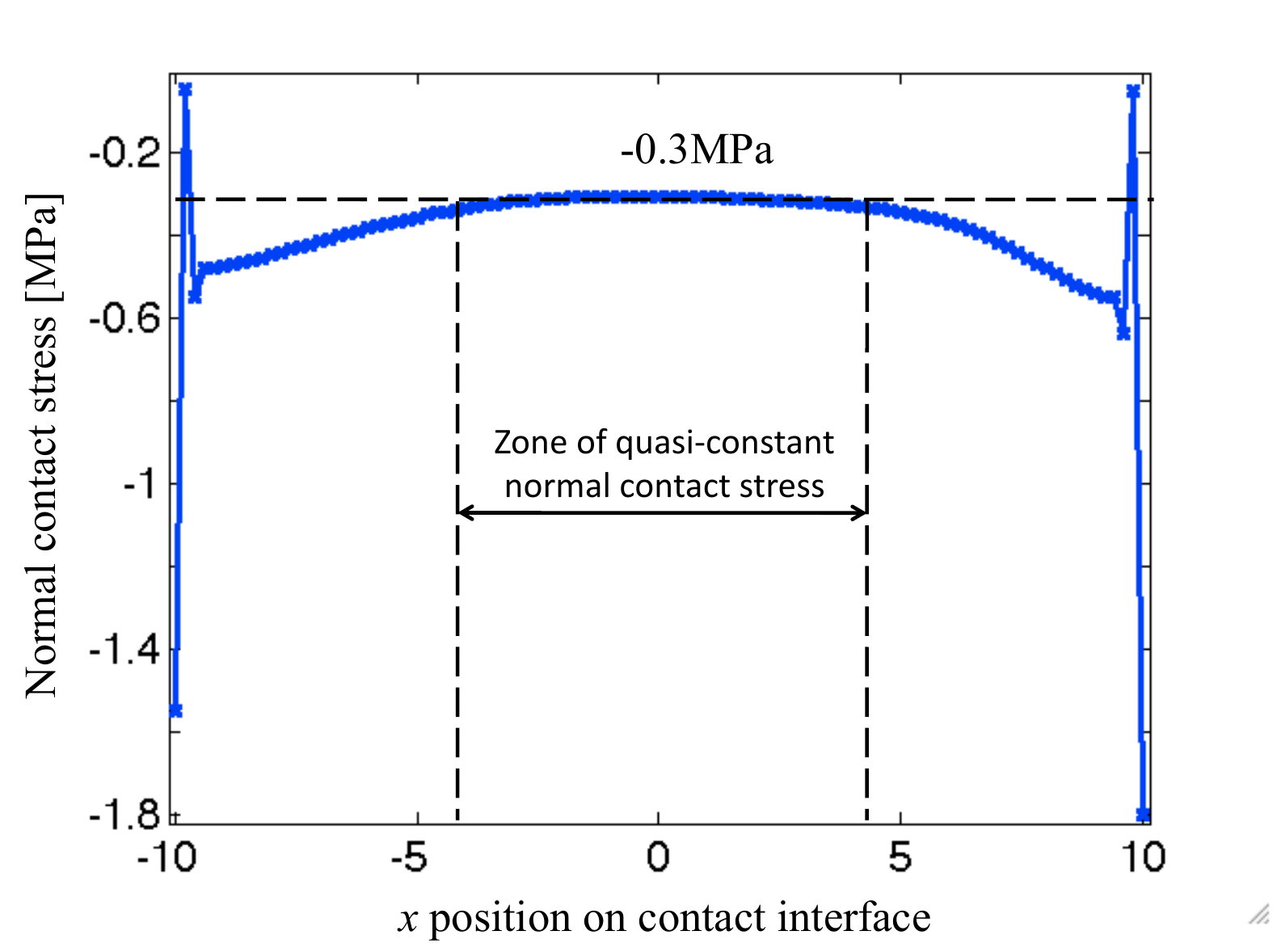}} 
\subfigure [ ]{\includegraphics[width = 3.in, height=3.in]{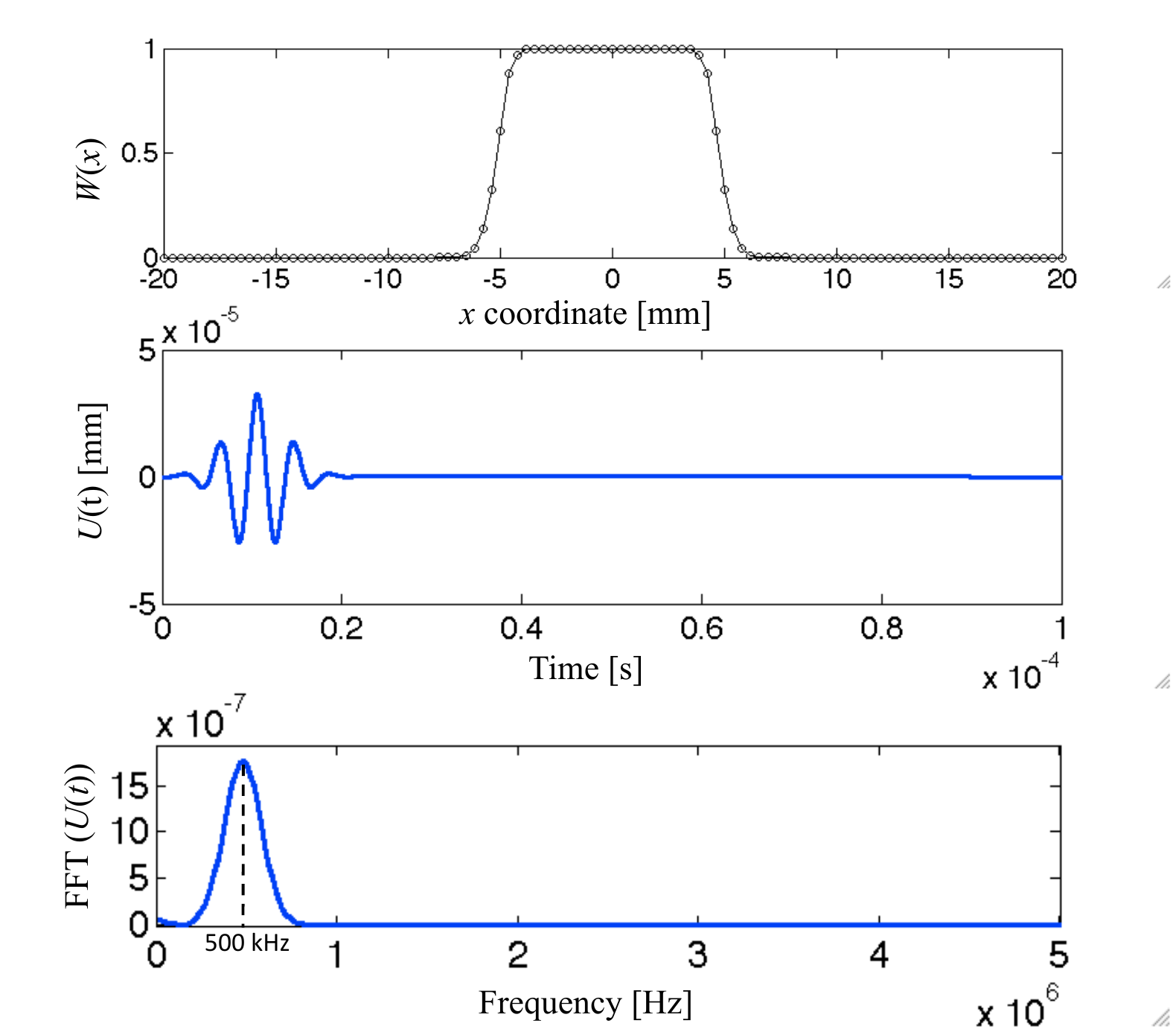}} 
\end{center} 
\caption{  (a) Normal contact stress along contact interface after the first step of simulations for $|\sigma_{\!_N}^0|$ = 0.3MPa. (b) The  imposed displacement on the upper surface $u^{(1)}  = W(x)U(t)$.    }
 \label{fam2}
\end{figure}

In the first step of the simulation a normal contact stress between the two solids is applied by numerically imposing appropriate  boundary conditions on the external surfaces.  The lower surface of the body P$^{(2)}$  is fixed ($u^{(2)}  = 0$) and the upper surface of the body P$^{(1)}$  is shifted by the appropriate amount ${\pmb v}_0$ in the $y$ direction ($u^{(2)}  = {\pmb v}_0$). The normal contact stress profile is shown in Figure \ref{fam2}a. The normal contact stress $\sigma_{\!_N}$ is quasi-constant on a part of interface (in Figure \ref{fam2}a, $\sigma_{\!_N}^0$=-0.3MPa).

In the second step, a non-planar shear wave is generated by imposing a dynamic displacement along the $x$ axis on the upper surface of P$^{(1)}$. During this stage of the simulation the lower surface of body P$^{(2)}$ is fixed ($u^{(2)}  = 0$) and the displacement $u^{(1)}$  imposed  on the upper surface of P$^{(1)}$ is expressed as a product of a weight function $W(x)$ and a function of time $U(t)$  shown in Figure \ref{fam2}b. The interface is supposed to remain 
in contact during the transmission/reflection of the shear wave. This assumption has been verified for all simulations.

\subsection{Numerical model}

PLAST2, an explicit dynamic finite element code in 2D, is used to simulate the behaviour of the system (P$^{(1)}$  and P$^{(2)}$) during frictional contact. PLAST2 is designed for large deformations and non-linear material behaviour  \cite{MNS12}. It uses a forward Lagrange multiplier method for the contact between deformable bodies. For this dynamic study, the formulation is discretized spatially  using the finite element method and discretized temporally {by  the} $\beta_2$ method. The contact algorithm uses slave nodes (situated on the contact surface of P$^{(1)}$) and target surfaces (on the contact surface of P$^{(2)}$) described by a four node quadrilateral element with 2$\times$2 Gauss quadrature rule. The elementary target segments are described by two nodes and approximated by bicubic splines \cite{MNS13}.

The forward Lagrange multiplier method is formulated for the  equations of motion at time  
$t^i = i \Delta t$ with the displacement condition imposed on the slave node at time 
$t^{i+1}$:
\begin{equation}\label{001}
{\bf M}\ddot{\bf U}^i+{\bf C}\dot{\bf U}^i+{\bf K}{\bf U}^i + {{\bf G}^{i+1}}^T
\pmb{\lambda}^i = {\bf F} ,
\quad
{\bf G}^{i+1}{\bf U}^{i+1} \le 0,
\end{equation}
where ${\bf M}$, ${\bf C}$ and ${\bf K}$ are  symmetric and positive  definite  matrices of mass, Rayleigh's proportional damping $({\bf C}=d_1{\bf M}+d_2{\bf K} )$ and stiffness of the system, respectively. 
${\bf U}$, $\dot{\bf U}$, $\ddot{\bf U}$ are respectively the vectors of nodal displacements, nodal velocities and nodal accelerations.   ${\bf F}$ is the vector of nodal external forces.  The vector 
$\pmb{\lambda} = [ \pmb{\lambda}_N, \pmb{\lambda}_T]^T$ comprises normal (N) and tangential (T) forces at contact points $C_m$. 
${\bf G}^T = [{\bf G}_N^T, {\bf G}_T^T]$ is the global matrix of the displacement conditions ensuring non-penetration and the contact law of the bodies in contact.

\begin{figure}[htb]  
\begin{center} 
\subfigure[] {\includegraphics[width=2.2in]{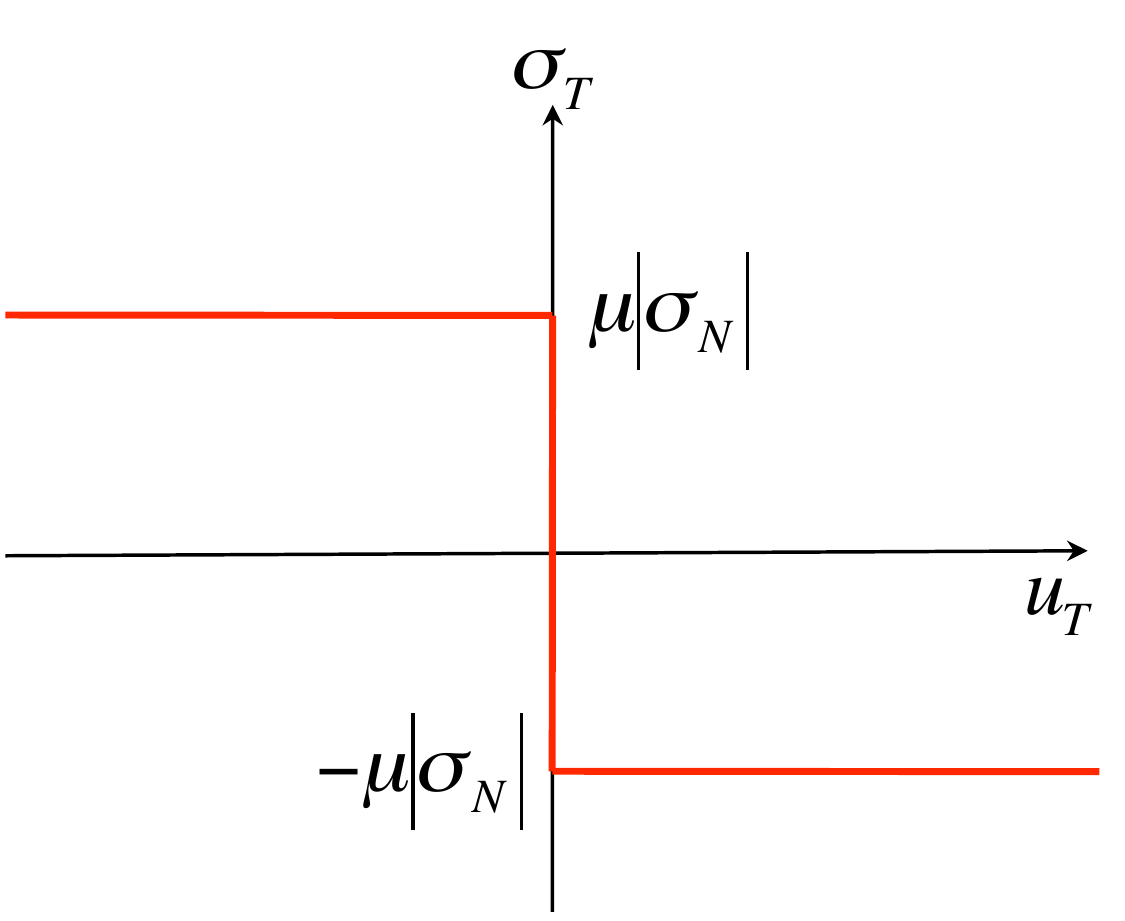}} 
\subfigure []{\includegraphics[width=2.4in]{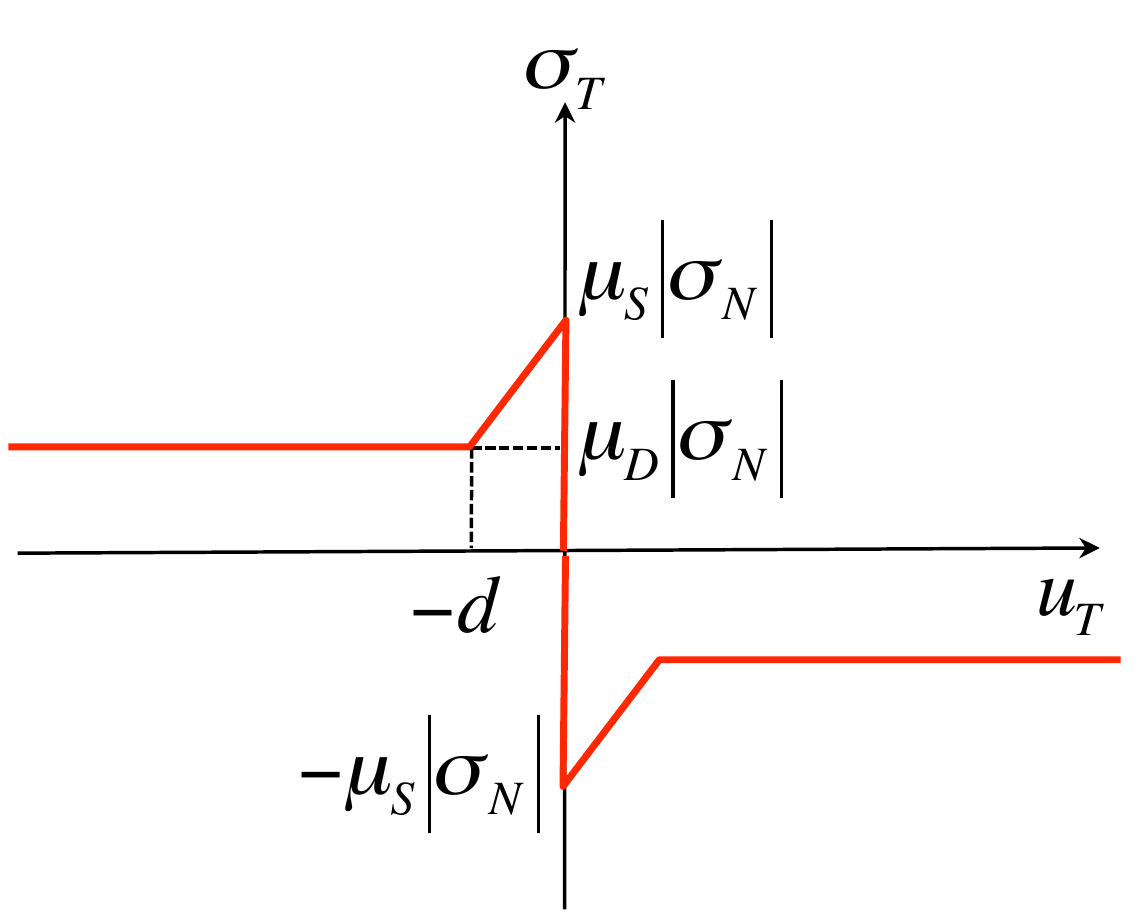}} 
\end{center} 
\caption{Tangential stress as a function of slip for Coulomb friction law  (a) and for slip-weakening friction law (b).    }
\label{fam4}
\end{figure}

The equations of motion \eqref{001} are discretized in time  using an explicit Newmark scheme. The {vectors     are} expressed at each time step using a time scheme of type 
$\beta_2$, $(\beta_2\in [0.5,1[)$: 
\begin{equation}\label{002}
\begin{cases}
\ddot{\bf U}^i &= \frac 2{{\Delta t}^2} \big( {\bf U}^{i+1}- {\bf U}^i - \Delta t
\dot{\bf U}^i\big),
\\
\dot{\bf U}^i  &= \frac 1{1+2\beta_2} 
 \big( \dot{\bf U}^{i+1} +\Delta t (1-\beta_2) \dot{\bf U}^{i-1}
+ \frac {2\beta_2} {{\Delta t}} ( {\bf U}^{i+1}- {\bf U}^i )\big).
\end{cases}
\end{equation}
The displacements $^*{\bf U}^i$   of the nodes situated on the contact surface (P$^{(1)}$ and P$^{(2)}$) are first computed with $\pmb{\lambda}^i$  equal to zero. For simplicity   $\beta_2$ is fixed at 0.5 and the nodal displacements $^*{\bf U}^{i+1}$ at time $t^{i+1}$  are obtained according to 
\begin{equation}\label{003}
^*\ddot{\bf U}^{i+1} =  {\Delta t}^2  {\bf M}^{-1} ({\bf F}- {\bf K}{\bf U}^i ) + 2 {\bf U}^i - {\bf U}^{i-1} .
\end{equation}
A constraint matrix ${\bf G}^{i+1}$  is formulated for the slave nodes if they have penetrated through a target segment. Calculation of the contact forces  $\pmb{\lambda}^i$  and of the correct nodal displacements ${\bf U}^{i+1}$  at time  $t^{i+1}$   are then performed:
\begin{equation}\label{004}
\begin{cases}
\pmb{\lambda}^i  = \big(  {\Delta t}^2   {{\bf G}^{i+1}} {\bf M}^{-1} {{\bf G}^{i+1}}^T\big)^{-1}
\, {{\bf G}^{i+1}} \, {^*{\bf U}}^{i+1}, 
\\
{\bf U}^{i+1} = {^*{\bf U}}^{i+1} -  {\Delta t}^2 {\bf M}^{-1} {{\bf G}^{i+1}}^T
\pmb{\lambda}^i .
\end{cases}
\end{equation}
Equations \eqref{004} are solved using the Gauss-Seidel method. The contact conditions solved during each internal iteration of this method are expressed as follows :


\begin{align}\label{005}
&\sigma_{\!_N} \le 0, \quad (\text{contact if }\sigma_{\!_N} \le 0 \text{ and separation if } \sigma_{\!_N}=0)
\nonumber \\
& |\sigma_{\!_T}| \le \mu_a | \sigma_{\!_N} | , \  \text{ and }
\begin{cases}
{v}_{\!_T} = 0&    \text{ if }  |\sigma_{\!_T} | <\mu_a |\sigma_{\!_N}| \ \  \text{ (stick)},
\\
\sigma_{\!_T} { v}_{\!_T} \le  0 &  \text{ if } |\sigma_{\!_T}| = \mu_a |\sigma_{\!_N}| \ \  \text{ (slip)},
\end{cases}
\end{align}
 \begin{figure}[H]  
\begin{center} 
\subfigure[ ] {\includegraphics[width=2.8in]{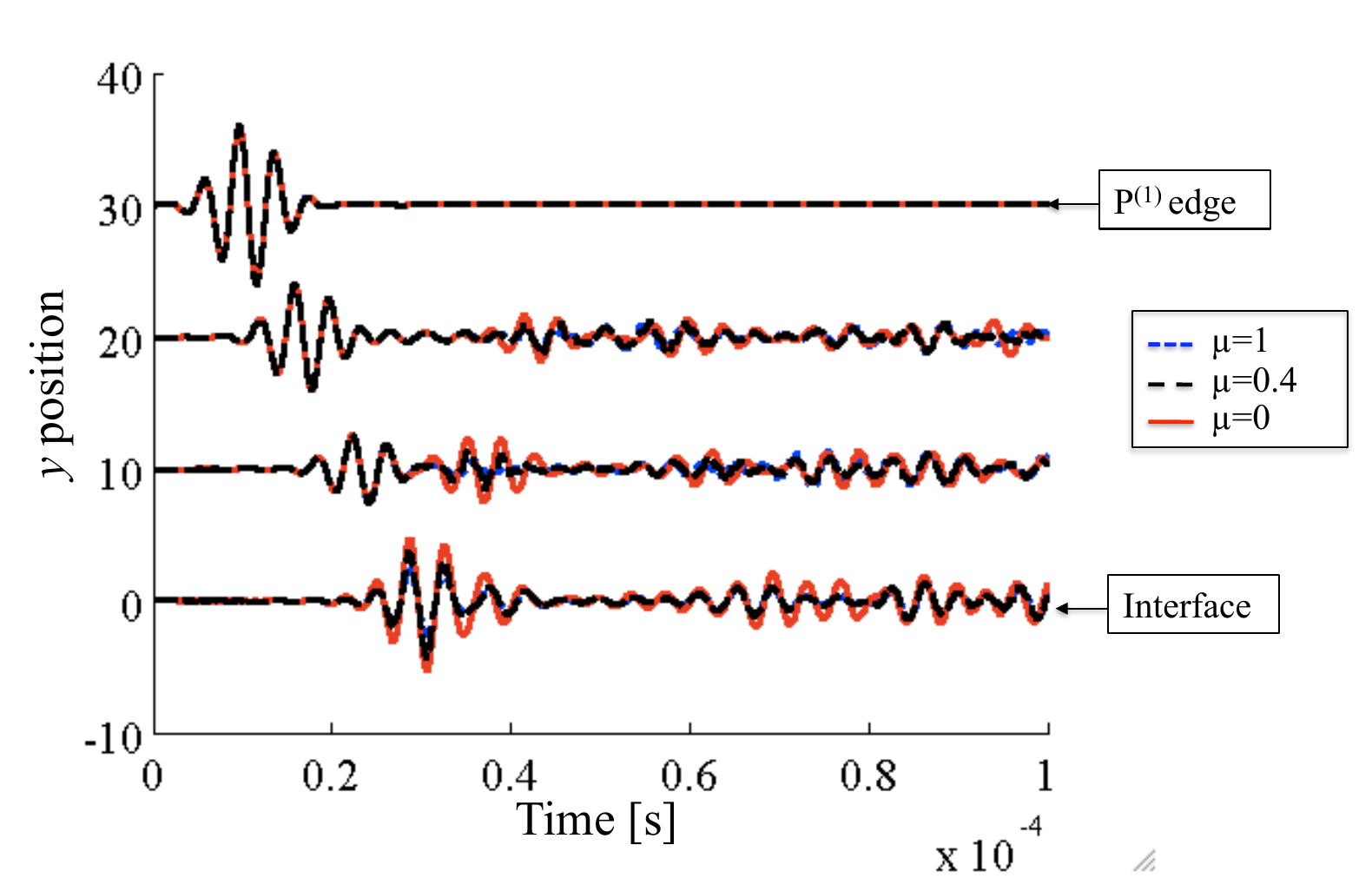}} 
\subfigure [ ]{\includegraphics[width=2.8in]{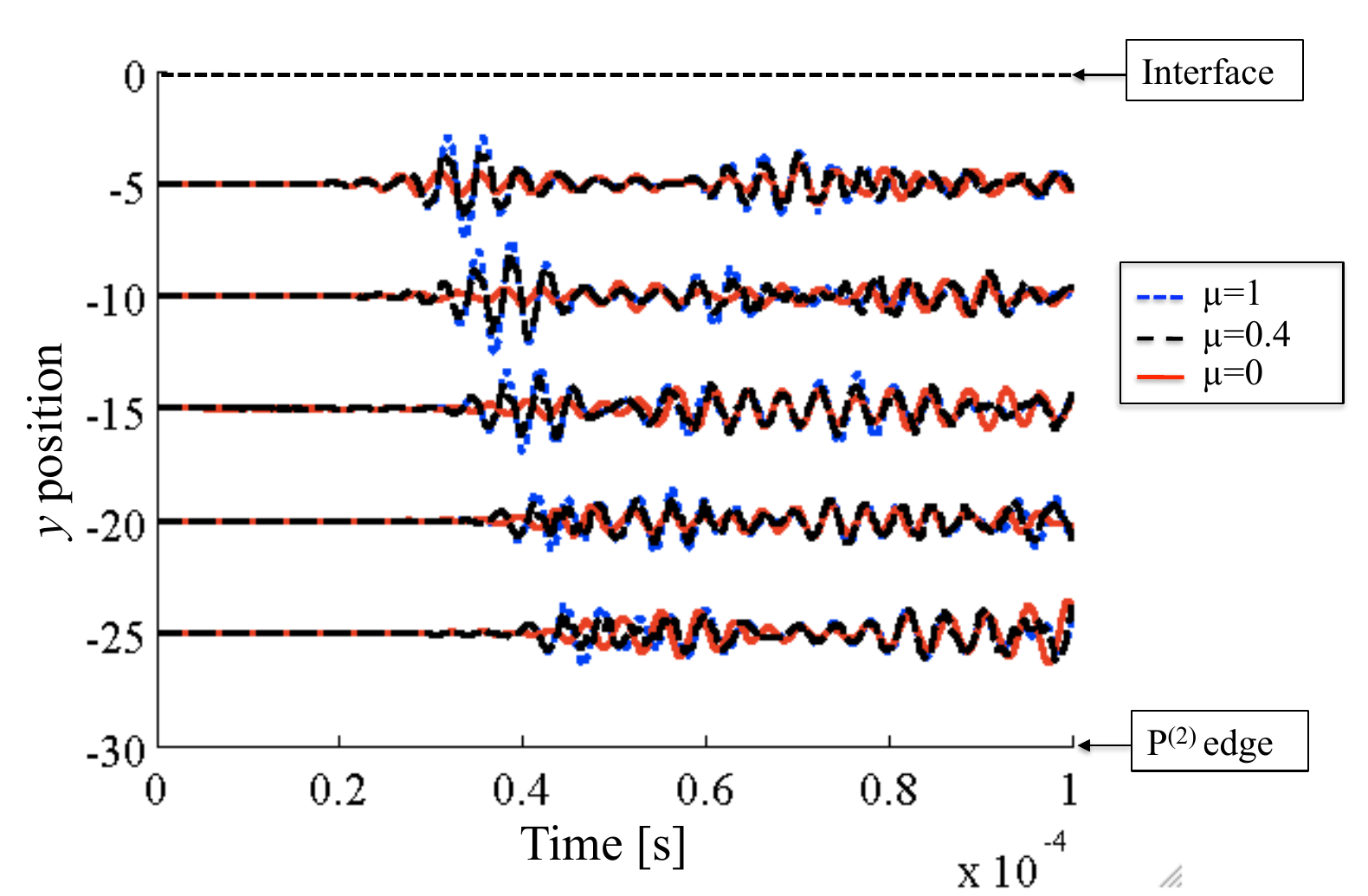}} 
\subfigure [ ]{\includegraphics[width=2.8in]{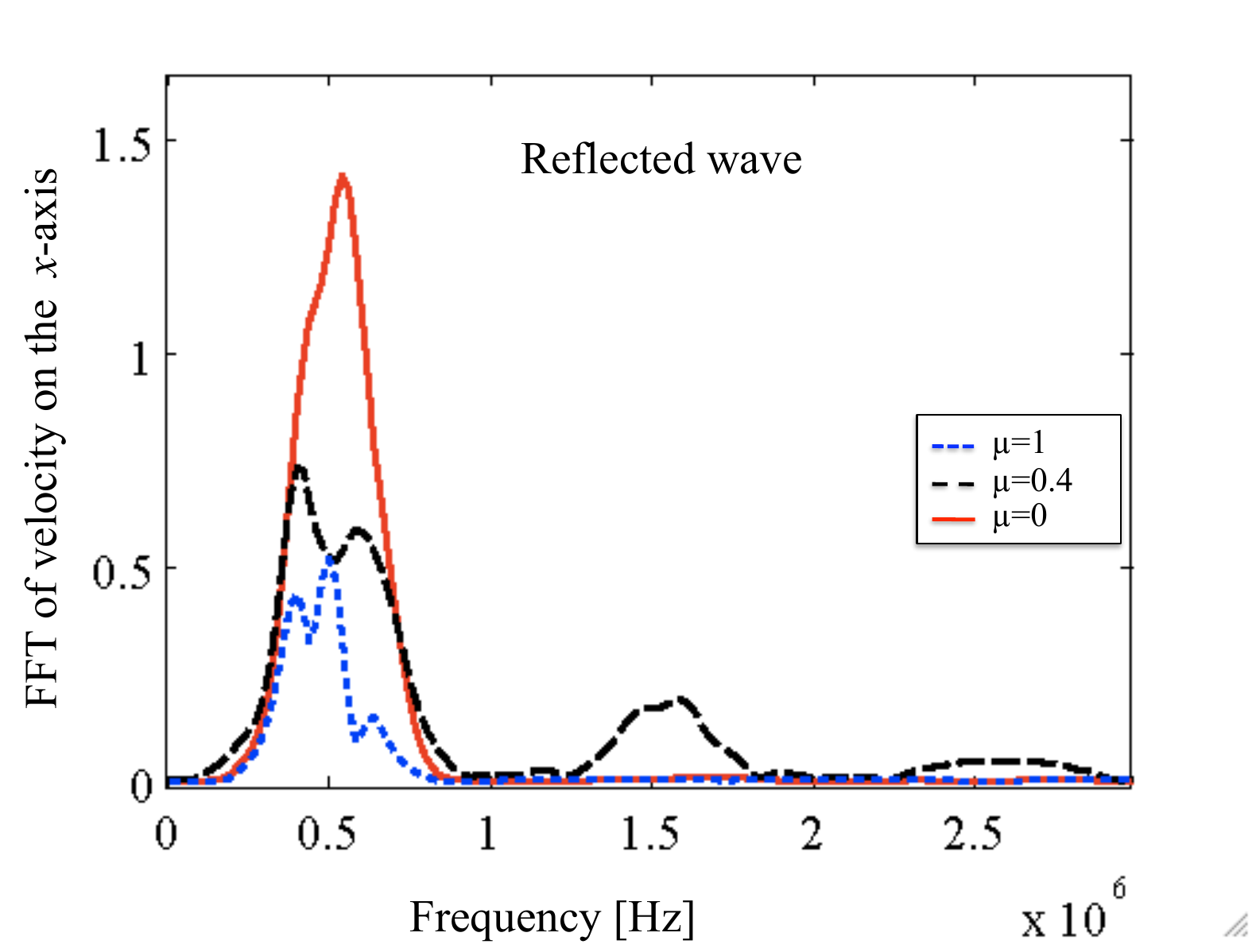}} 
\subfigure [ ]{\includegraphics[width=2.8in]{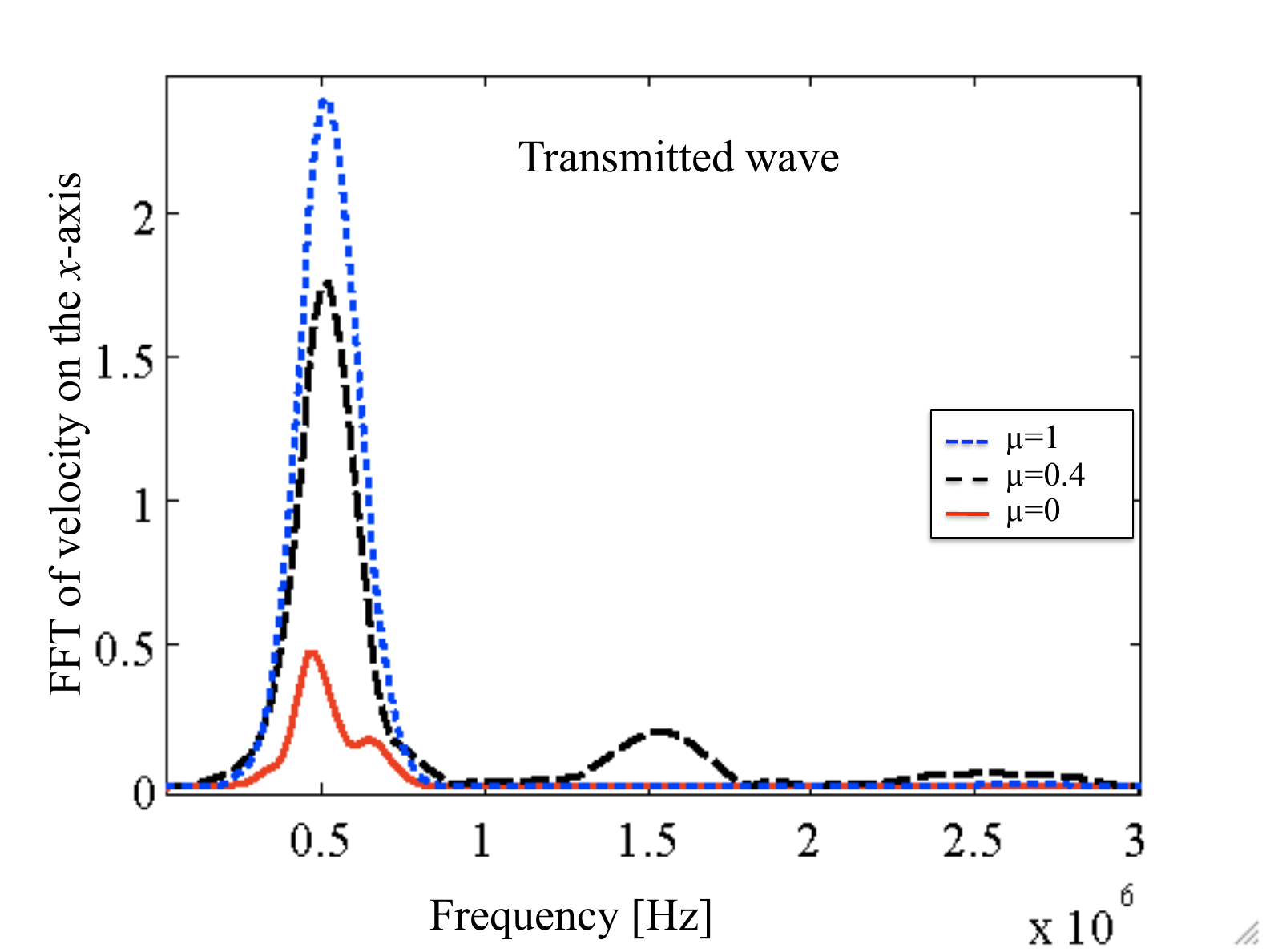}} 
\subfigure [ ]{\includegraphics[width=3in]{amf4c}} 
\end{center} 
\caption{Propagation of shear wave in (a) the   body P$^{(1)}$ and   (b) in P$^{(2)}$, both for $x=0$.  The waterfall plots show  the $x$-velocity versus time for three different friction coefficients ($\mu$= 0, 0.4 and 1).
 FFT of $x$-velocity of (c) reflected  and (d) transmitted waves for three different friction coefficients   ($\mu$= 0, 0.4 and 1). The FFT processing of the reflected and transmitted wave are respectively done at point A ($x=0$,$y=20$mm) and point B ($x=0$,$y=-5$mm).  }
\label{fam5}
\end{figure}

\noindent
where ${v}_{\!_T}$  is the relative tangential velocity for a slave node related to the  surface, $\sigma_{\!_N}$ and $\sigma_{\!_T}$ are normal and tangential stress respectively, subscript ${N}$  and ${T}$ referring to the normal and tangential vectors defining the contact. $\mu_a$ is an apparent friction coefficient.
Two friction laws are considered in the simulations : a standard Coulomb friction law with a constant friction coefficient ($\mu_a$=$\mu$, see Figure \ref{fam4}a) and a slip-weakening (SW) friction law (Figure \ref{fam4}b) for which $\mu_a$ is defined in  \cite{MNS14,MNS15} as:  
\begin{equation}\label{0051}
\mu_a(u_{\!_T})=
\ \ 
\begin{cases}
\mu_S-\frac {(\mu_S-\mu_D)} {{d}} u_{\!_T} , &	u_{\!_T} \le d,
\\
\mu_D, &	u_{\!_T} > d,
\end{cases}
\end{equation}
where $\mu_S$ and $\mu_D$ are static and dynamic friction coefficient, $u_{\!_T}$ is the slip and $d$ the critical slip distance. Once the dynamic regime is reached, the friction corresponds to a classical Coulomb friction law with a dynamic friction coefficient $\mu_D\ (< \mu_S)$. This physically motivated model of slip weakening friction (i.e. the decrease of the friction force with slip), which was first introduced in the geophysical {context \cite{MNS17}, can} be associated to a phenomenon of adhesion caused by long time stationary contact \cite{MNS16}. For more details on rate and state-variable friction law (including the slip-weakening one) see for example \cite{MNS18}.

\subsection{Results}\label{3.3}
\subsubsection{Case of Coulomb friction law}

Figure \ref{fam5} shows the velocity on the $x$-axis versus time and $y$-position (depth) for three different friction coefficients. At time $t = 0$, the shear wave is generated on the upper face of P$^{(1)}${(see Fig. \ref{fam1})}.  The incident wave then propagates through P$^{(1)}$ and the  contact interface gives rise to reflected and transmitted waves. 

It  can be seen from Figure \ref{fam5} that the magnitude of the transmitted wave is increasing with the friction coefficient.  Consequently, the magnitude of the reflected wave is decreasing. Points A ($x$=0,$y$ =20mm) and B ($x$=0,$y$=-5mm) are chosen {for   Fast} Fourier Transform (FFT) processing of the reflected and transmitted waves,  respectively.

Figure \ref{fam5}c and  \ref{fam5}d shows the FFT of velocity along the $x$-axis for the reflected and transmitted wave for $\mu$=0, $\mu$=0.4 and $\mu$=1. Nonlinear harmonics should appear in the spectrum as soon as slipping zones occur at the contact interface. For friction coefficient equal to 0 or 1, no harmonics are observed in the reflected and transmitted wave spectrum. For the case $\mu$= 0, the interface is in a sliding state during transmission of shear waves but  the system is perfectly linear under the previously made assumption of no separation. When the friction coefficient is equal to 1, the contact law is nonlinear, but the friction coefficient is too large  to permit enough sliding such that would produce a noticeable effect on the transmitted/reflected wave spectrum. In other words, during transmission/reflection of the  shear wave the contact nodes remain almost always in the sticking state of the Coulomb friction law. For friction coefficient equal to 0.4, the third harmonic (1.5MHz) is observed in the reflected and transmitted wave spectra due to the nonlinearity of the friction law and the occurrence of sliding zones  during the transmission/reflection stage.

As expected,  the magnitude of the wave  decreases along the propagation direction because of diffraction effects.  In order to compare the numerical results with the analytical ones presented in \S \ref{sec2}, the non-dimensional parameter $\xi= \mu |\sigma_{\!_N}|/\sigma_{\!_{TM}}$ used in the analytical model (where $\sigma_{\!_{TM}}$ denotes the maximal  stress magnitude for  the  plane wave) needs to be modified. Let $\xi=1$ correspond to the case in which the wave magnitude is not large  enough to initiate  slip at the interface. From this definition a critical friction coefficient for slip, $\mu_{crit}$, is  determined  numerically, and the modified $\xi_{num}$ is  defined as follows:
\begin{equation}\label{3=0}
\xi_{num} = \frac{\mu |\sigma_{\!_N}^0|}{\sigma_{num}}, \quad  \text{with  }
\sigma_{num}  = \mu_{crit} |\sigma_{\!_N}^0|.
\end{equation}
The harmonic magnitudes  ${A}_{i}^v $ are extracted from the FFT of the computed velocity (Figure \ref{fam5}c and d). Energy dissipated during transmission of the shear wave $D(x)$ is numerically calculated along the interface from its definition \eqref{7}$_1$.  Since the incident wave is non-plane, $D(x)$ is not constant along the interface, and accordingly a mean value $D$ on the zone of constant normal stress (Figure \ref{fam2}a) is chosen for the results. 
Non-dimensional measures of the harmonic magnitudes and  energy dissipated per unit cycle during  transmission of the shear wave are defined respectively as 
\begin{equation}
\overline{A}_i = \frac{A_{i}^v}{A_{1_{max}}^v },
\quad
\overline{D}  = \frac{\omega Z}{ 5\sigma_{num}^2 } D,
\label{30-=}
\end{equation}
where $A_{1_{max}}^v$ is  the magnitude of the first harmonic of the transmitted wave velocity for $\xi=1$ at the probe point B and the factor $5$ is the number of cycles in the incident signal (see Figure \ref{fam2}b for $U(t)$).
\begin{figure}[H]  
\begin{center} 
\subfigure[] {\includegraphics[width=3.in, height=2.7in]{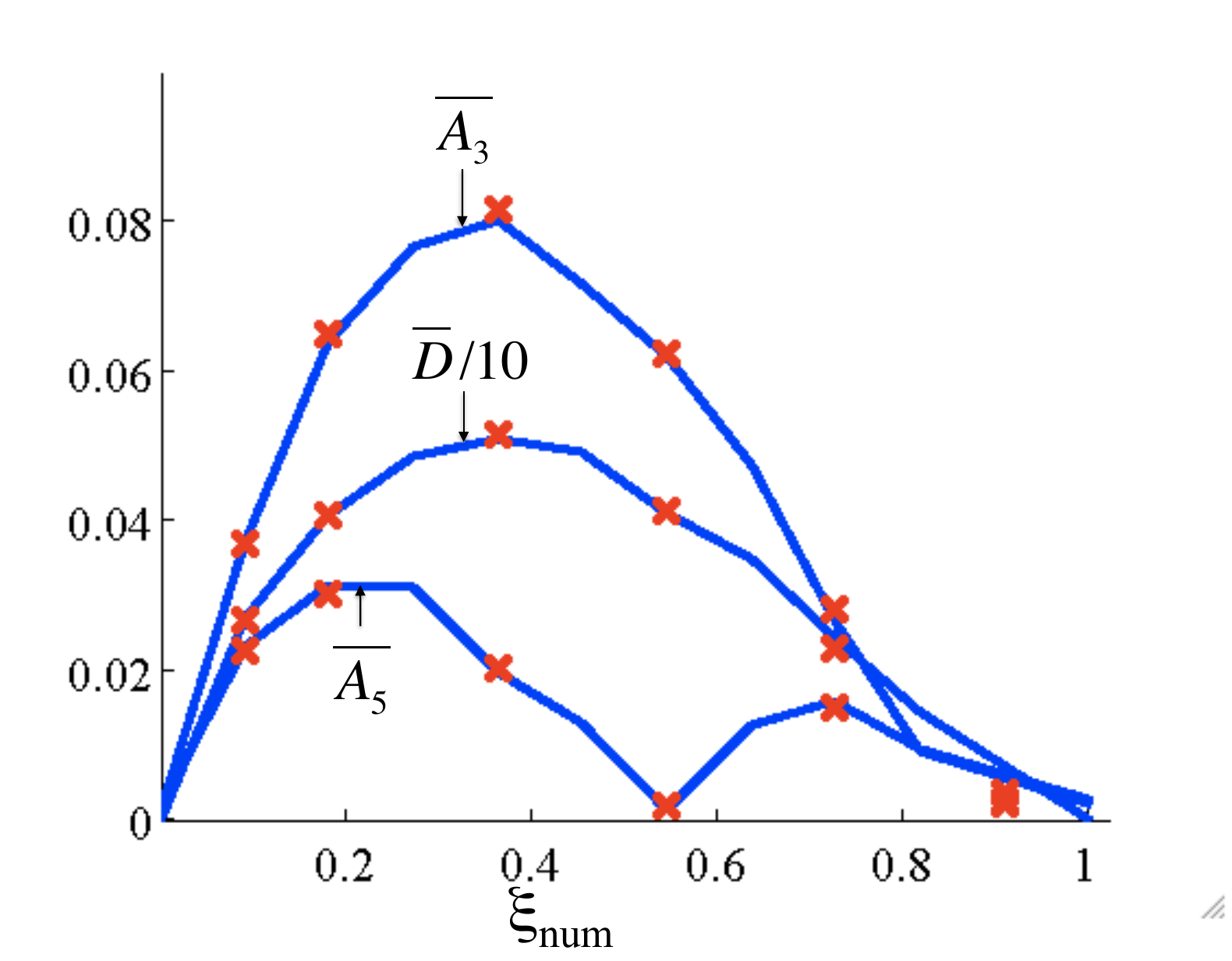}}
\subfigure[] {\includegraphics[width=3.in, height=2.7in]{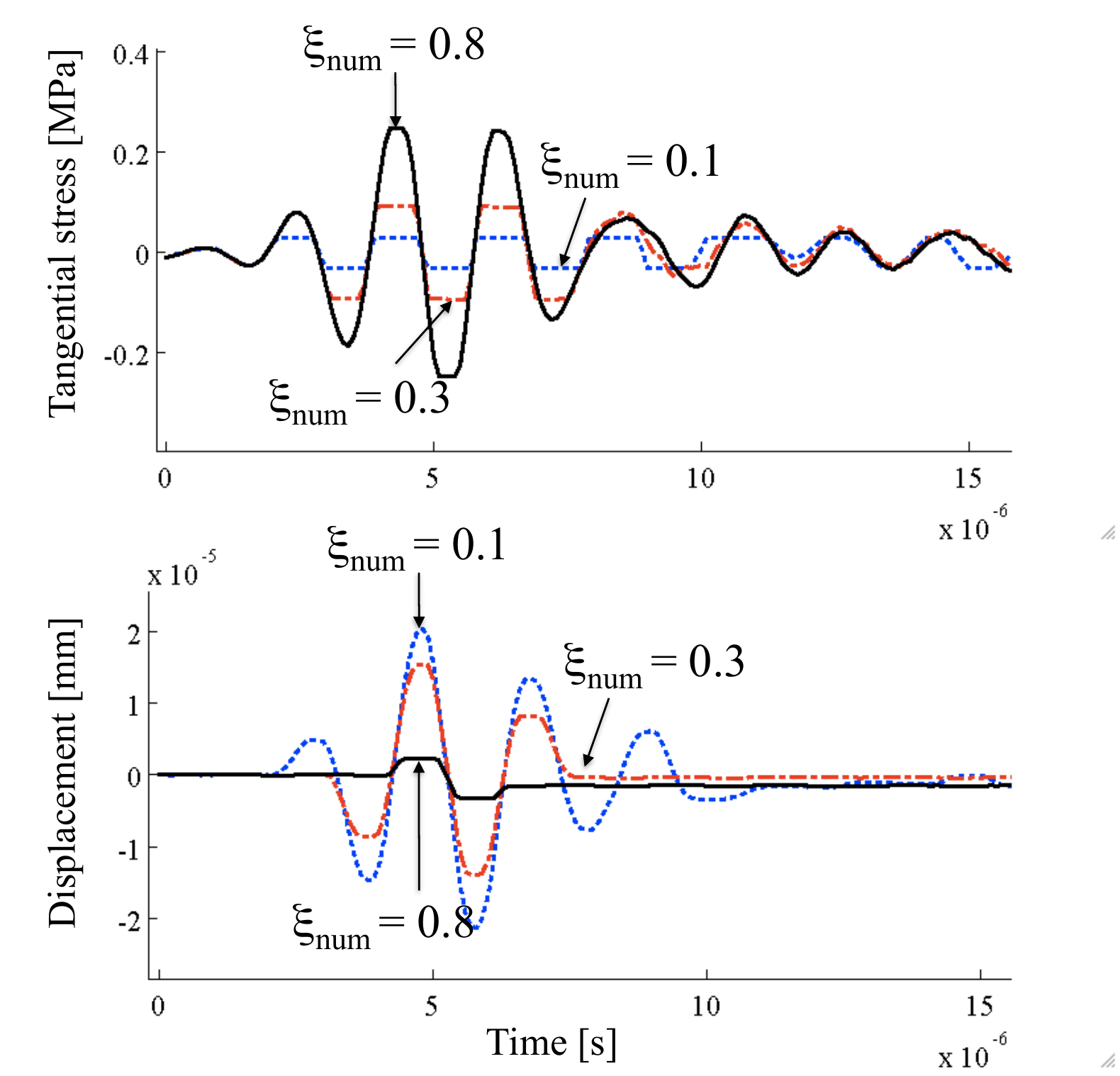}}
\end{center} 
\caption{(a) Third and fifth harmonics $\overline{A}_3$, $\overline{A}_5$ and dissipated energy $\overline{D} $ for $|\sigma_{\!_N}^0|$ = 0.3MPa (--------) and for $|\sigma_{\!_N}^0|$ = 0.6MPa ($\times$). 
(b)  Tangential stress and relative displacement on interface ($x$=0)  versus time for three values of $\xi_{num}$.  The results are presented for  the case $|\sigma_{\!_N}^0|$=0.3MPa.
}
\label{fam7}   
\end{figure}
  
Figure \ref{fam7}a displays the evolution of the non-dimensional third and fifth harmonic magnitudes and  non-dimensional dissipated energy versus $\xi_{num}$ for two different values of initial normal stress: 0.3MPa and 0.6MPa. It can be seen that the initial normal contact stress has no influence on the non-dimensional third, fifth harmonics and dissipated energy. 
Comparing Figures \ref{fam7}a and \ref{fig5}  shows a good qualitative agreement between the numerical and analytical results for $\overline{D} $ and $\overline{A}_i $. The fact that the values obtained numerically are about {half of}  the analytical ones is because the sliding is not uniform for a cylindrical shape of wave front used in the numerical model, while the analytical model assumes a plane wave.

The appearance  of odd harmonics in the spectrum is due to distortion of the incident wave during the transmission/reflection as the interface  partially slides under friction. The evolution of the odd harmonics, directly related to this distortion, depends on two competing  effects. 
First, for a fixed value of the sliding distance,  more energy is dissipated and  the wave is  more highly distorted as the friction coefficient increases. On the other hand, since the incident wave is the same for the different simulations, the  cumulative sliding distance is actually diminished as the friction coefficient increases, as shown in Figure \ref{fam7}b. These two antagonist aspects of nonlinearity of a frictional interface explain the existence of a distinguished value of the normal stress that maximizes energy dissipation:  $\xi_{num} \approx$   0.4 in Figure \ref{fam7}a.

The numerical simulations allow us to explore dependence on parameter values describing  the contact interface (displacement, normal and tangential stresses, sliding distance,...). Figure \ref{fam7}b shows the tangential stress and displacement  of the contact node situated on the interface at  $x$ = 0 mm  versus time. The displacement  corresponds to the tangential displacement of the slave node (of P$^{(1)}$) in a local frame of  P$^{(2)}$ whose origin is situated at the initial contact position of this node (of P$^{(1)}$) on the P$^{(2)}$ interface before the arrival of the incident shear wave. As the friction coefficient   increases the relative displacement   decreases, because  the resistance to sliding is increasing. For $\xi_{num}$ =0.1, the sliding of the slave node is important but the sinusoidal form is weakly affected, whereas for $\xi_{num}$ =0.8  the sliding of the slave node is weak, but the evolution of the displacement of the contact node is strongly distorted from the initial waveform.

\subsubsection{Case of slip-weakening (SW) friction law}

In this {simulation} the numerical model takes into account a slip-weakening friction {law that includes \cite{MNS14,MNS15}}  the combined effects of static and dynamic friction, with coefficients 
$\mu_S$ and $\mu_D$ respectively. A third parameter $d$ describes the slope of the decreasing  friction coefficient, see Figure \ref{fam4}b. For all simulations, static friction coefficient  and incident shear wave magnitude remain the same ( $\mu_S$=0.3 and $\sigma_{\!_{TM}}$=0.33MPa). The simulations  were performed for different values of the initial normal stress $|\sigma_{\!_N}^0|$ (from 0.15 MPa to 1.2 MPa). Note that in these conditions no slip is allowed when $|\sigma_{\!_N}^0|$ is greater than 1.1MPa.  

\begin{figure}[htb]  
\begin{center} 
\subfigure[] {\includegraphics[width=3.in, height=3.2in]{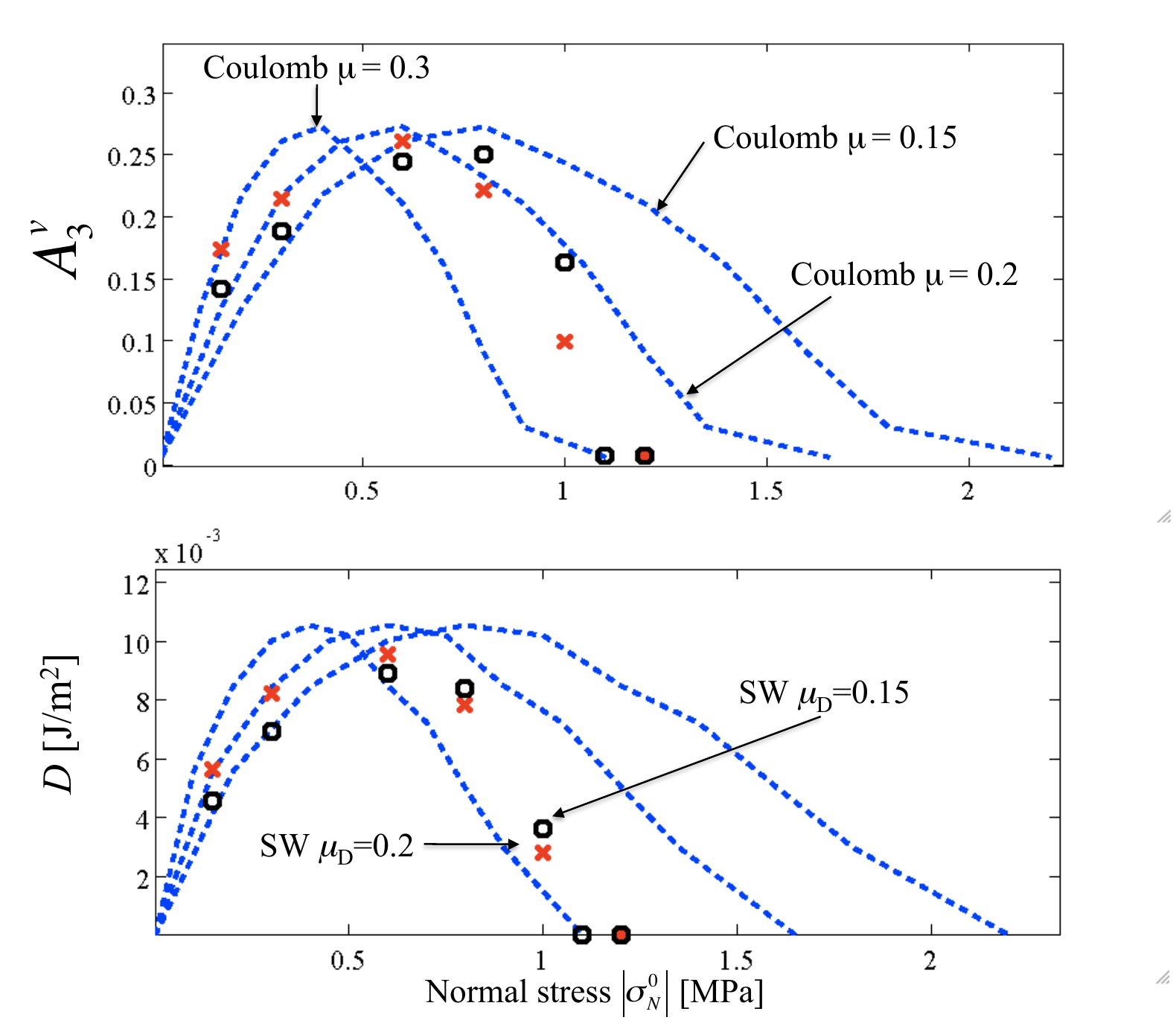}}
\subfigure[] {\includegraphics[width=3.in, height=3.2in]{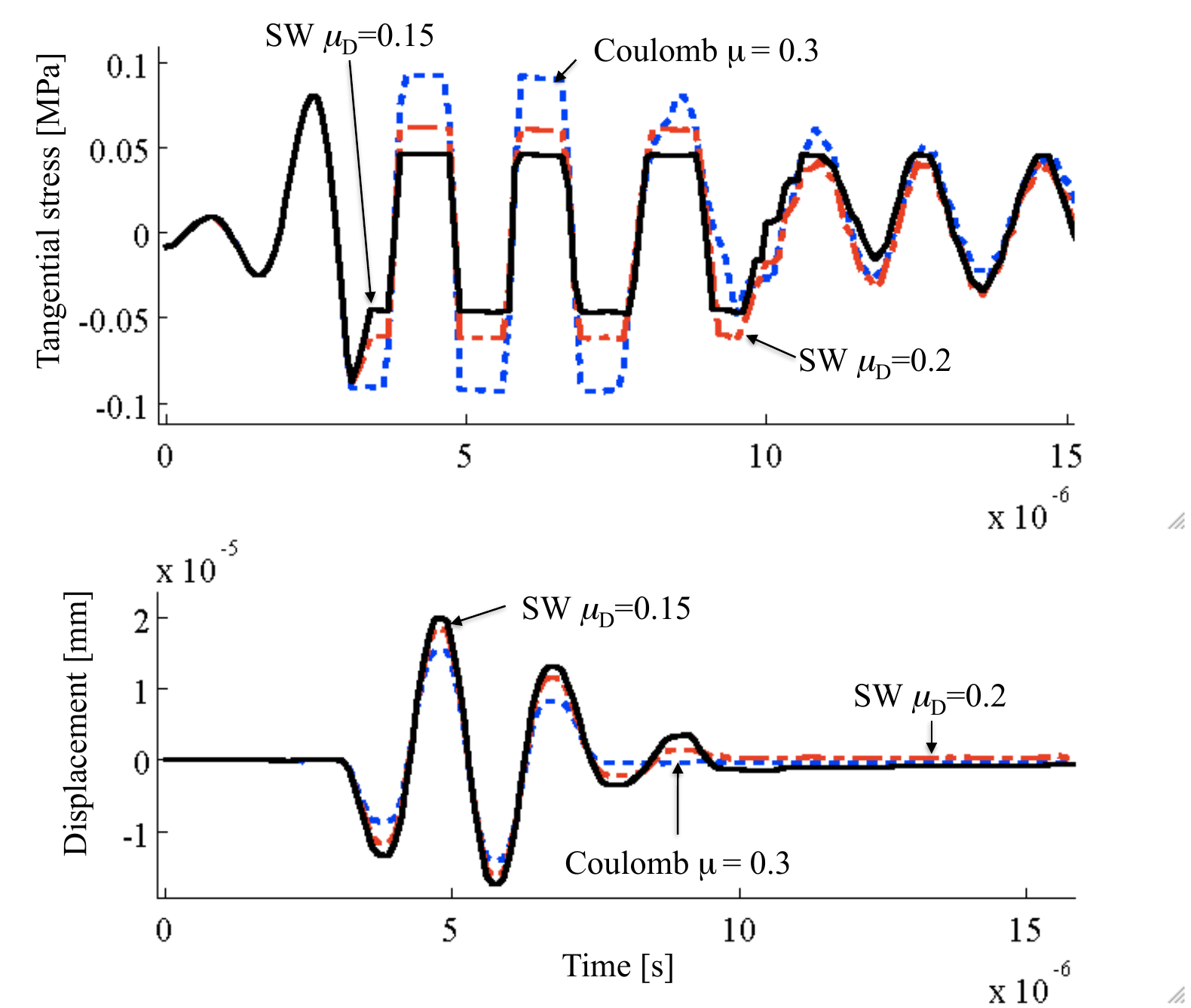}}
\end{center} 
\caption{(a) Evolution of the third harmonic $A_{3}^v$ and of the dissipated energy per unit surface ${D}$ as a function of  $|\sigma_{\!_N}^0|$ for Coulomb and SW laws for different friction coefficients ($\mu_S=$ 0.3, $d = 10^{-2} \mu$m). 
(b) Tangential stress and displacement on the interface ($x=0$) versus time for  the case $|\sigma_{\!_N}^0|=0.3$MPa.
}
\label{fam9}
\end{figure}

Figure \ref{fam9} shows the results for two dynamic friction coefficients in  comparison with those for the constant friction law. Figure \ref{fam9}a confirms that for this model  the third harmonic again follows the behavior of the dissipated energy. Maximum values of the third harmonic $A_{3}^v$ and dissipated energy per unit surface $D$ are decreasing when the dynamic friction coefficient is decreasing. The maximum of both is obtained for a larger value of  $|\sigma_{\!_N}^0|$. 

Figure \ref{fam9}b shows the tangential stress $\sigma_{\!_{T}}$ and the tangential slip $u_{\!_{T}}$ of the contact node situated at  $x$ = 0 on the interface versus time. The interface is sticking because $|\sigma_{\!_{T}}|$ is less than $\mu_S |\sigma_{\!_N}^0|$. When $|\sigma_{\!_{T}}|$ is equal to $\mu_S |\sigma_{\!_N}^0|$, the system begin to slide. $|\sigma_{\!_{T}}|$ remains the same for constant Coulomb friction ($\mu =\mu_S$) while, for the SW law,  $|\sigma_{\!_{T}}| = \mu_D|\sigma_{\!_N}^0|$ after the slip distance becomes greater than $d$. Hereafter the interface behavior corresponds to a classical Coulomb friction law with $\mu =\mu_D$.
As expected, since the dynamic friction coefficient is less than the static one, the relative displacement at the interface is greater than for the constant friction coefficient case ($\mu =\mu_S$). Also,  the relative displacement   increases when the dynamic friction coefficient decreases. The resistance to  motion due to friction is less important and so  the dissipated energy is reduced. This explains the offset of the position of the maximum dissipated energy $D$ (Figure \ref{fam9}a). The form of the curve ${A}_3 (|\sigma_{\!_N}^0|)$ is directly related to  the interface friction law in effect. 

\begin{figure}[htb]  
\begin{center} 
\subfigure[] {\includegraphics[width=3.in, height=2.7in]{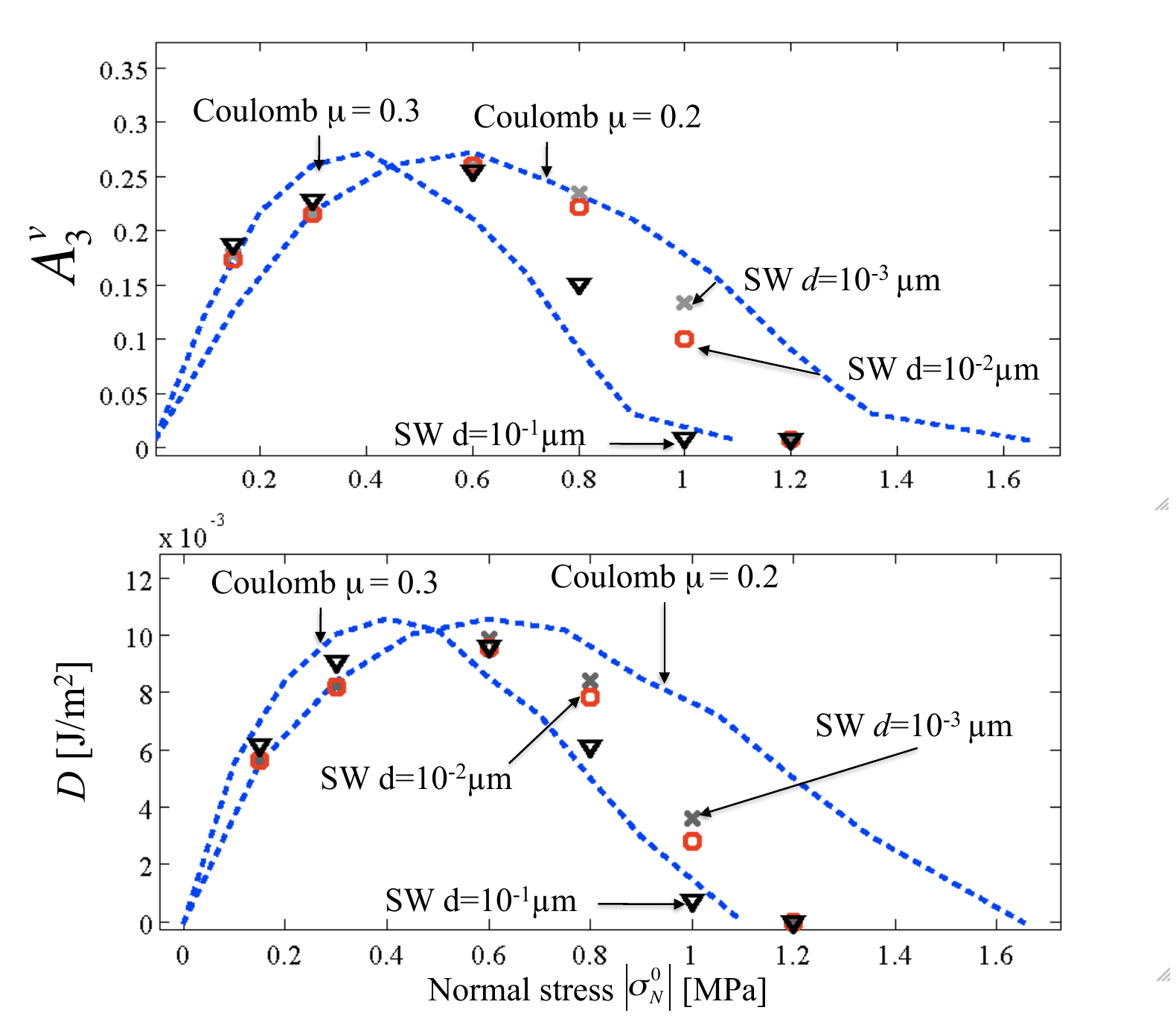}}
\subfigure[] {\includegraphics[width=3.in, height=2.7in]{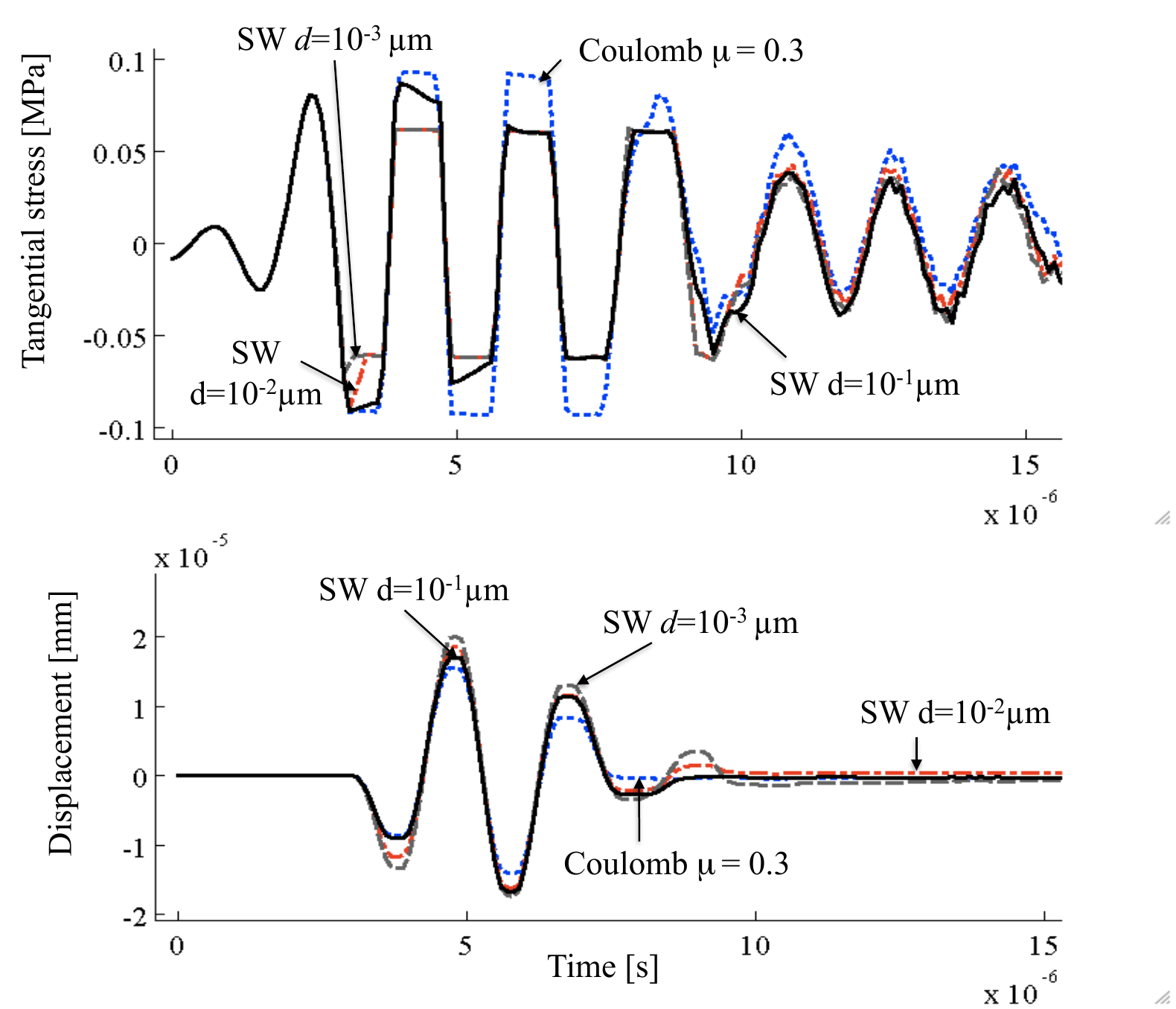}}
\end{center} 
\caption{(a) Evolution of the third harmonic $A_{3}^v$ and of the dissipated energy per unit surface ${D}$ as a function of  $|\sigma_{\!_N}^0|$ for Coulomb ($\mu = 0.3$ and $\mu = 0.2$) and SW laws for different values of parameter $d$ ($\mu_S=0.3$, $\mu_D=0.2$). 
(b) Tangential stress and displacement on the interface ($x$=0) versus time for  the case $|\sigma_{\!_N}^0|=0.3$MPa.
}
\label{fam10}
\end{figure}

Figure  \ref{fam10} highlights an important effect of {the} parameter $d$ on the tangential stress and slip at the interface, and consequently on dissipated energy and third harmonic magnitude. For a low value of $d$ ($=10^{-3}\mu$m), the system behavior is similar {to the} constant friction law with $\mu = \mu_D$. The sliding is delayed due to the static friction coefficient. For low value of $|\sigma_{\!_N}^0| (< 0.6MPa) $  this delay has little influence on the interface behavior. On the contrary, if $|\sigma_{\!_N}^0| \ge$ 0.6MPa, the delay is growing and the interface slip decreases reaching zero for  $|\sigma_{\!_N}^0| $=1.1MPa. This limit value of $|\sigma_{\!_N}^0|$ depends on the parameter $d$ namely, it decreases as $d$ increases. As a result, for $d=10^{-1}\mu$m, the limit value of $|\sigma_{\!_N}^0| $ is very low and hence the dissipated energy and the third harmonic curves tends to those for constant friction coefficient $\mu =\mu_S$.

\section{Conclusions}\label{sec4}

This study presents a combined numerical and analytical analysis of the propagation of a shear wave through a frictional interface. A simple analytical model is developed that describes the case of a classical Coulomb friction law and a plane wave. The model pinpoints the pivotal non-dimensional parameters of the problem and yields an explicit dependence of the measurable quantities on these parameters. Good agreement between the analytical and numerical results for this case is demonstrated. The  numerical analysis is also applied to a more complex case which involves a non-planar wave and a slip-weakening (SW) friction law with static and dynamic coefficients $\mu_S$ / $\mu_D$. For the both cases, it is shown that the amplitude of the third harmonic is directly related to the friction-induced dissipated energy. The evolution of the third harmonic as well as of the dissipated energy depends on two aspects: on the cumulative sliding distance during the transmission of the shear wave and on the friction coefficient. When a SW friction law is taken into account, the maximum values of third harmonic magnitude and of the dissipated energy are achieved  for a larger  value of the applied normal stress $|\sigma_{\!_N}^0|$. This is explained by the fact that the static effect in the SW law delays the  interface sliding in comparison with the case $\mu= \mu_D$. Less sliding and $\mu_D< \mu_S$ both imply lower values of dissipated energy and of third harmonic.


The main conclusions of this study are as follows :\\
(i) There exists a specific value of the normal stress that maximizes energy dissipation. This value results from the competition of two antagonist aspects of a frictional interface: sliding (cumulative sliding distance) and resistance to motion (tangential stress).  \\
(ii) The SW friction law affects the form of the curves  {for the dissipated energy, $D(|\sigma_{\!_N}^0|)$, and third harmonic, } $A_3^v(|\sigma_{\!_N}^0|)$. The maximal value is obtained for a larger value of $|\sigma_{\!_N}^0|$. Each of the three parameters of {the SW friction model} has a different impact on interface sliding and dissipated energy: {the} dynamic friction coefficient  $\mu_D$ acts on the steady state sliding, {the} static friction coefficient $ \mu_S$ delays the sliding in comparison with the constant friction law  $\mu= \mu_D$, and $d$ acts directly on this delay. \\
(iii) For the form of shear incident wave used in this study, the third harmonic evolution follows the variation of the dissipated energy at the interface. Thus the measurement of transmitted or reflected wave signal can provide information on dissipated energy at the interface, and the form of the $A_3^v(|\sigma_{\!_N}^0|)$ curve can reveal the type of friction law of the interface.


\end{document}